\documentclass[showpacs,aps,twocolumn,superscriptaddress,floatfix,nofootinbib]{revtex4}
\usepackage{amsmath}
\usepackage[dvips]{graphicx,epsfig,color}

\usepackage{color}

\definecolor{grey}{rgb}{0.93,0.93,0.93}
\definecolor{lred}{rgb}{1,0.4,0.4}
\newcommand{\PGRcomm}[1]{{#1}}
\newcommand{\PGRnew}[1]{{#1}}
\newcommand{\PKcomm}[1]{{#1}}

\begin{document}

\title{Variations on a theme by Skyrme}
\author{P. Kl\"upfel}
\affiliation{Institut f\"ur Theoretische Physik II, Universit\"at
  Erlangen-N\"urnberg, Staudtstrasse 7, D-91058 Erlangen, Germany}
\author{P.--G. Reinhard}
\affiliation{Institut f\"ur Theoretische Physik II, Universit\"at
  Erlangen-N\"urnberg, Staudtstrasse 7, D-91058 Erlangen, Germany}
\author{T. J. B\"urvenich}
\affiliation{Frankfurt Institute for Advanced Studies,
Tuth-Moufang-Str. 1, 60438 Frankfurt am Main, Germany}
\author{J. A. Maruhn}
\affiliation{Institut f\"ur Theoretische Physik, Universit\"at Frankfurt, 
Max-von-Laue-Str. 1, 60438 Frankfurt am Main, Germany}
\date{draft, 11. January 2007}
\begin{abstract}
We present a survey of the phenomenological adjustment of the
parameters of the Skyrme-Hartree-Fock (SHF) model for a
self-consistent description of nuclear structure and low-energy
excitations. A large sample of reliable input data from nuclear bulk
properties (energy, radii, surface thickness) is selected guided by
the criterion that ground-state correlations should remain small.
Least squares fitting techniques are used to determine the SHF
parameters which accommodate best the given input data. The question of
the predictive value of the adjustment is scrutinized by performing
systematic variations with respect to chosen nuclear matter properties
(incompressibility, effective mass, symmetry energy, and sum-rule
enhancement factor).  We find that the ground-state data, although
representing a large sample, leave a broad range of choices, i.e. a
broad range of nuclear matter properties. Information from giant
resonances is added to pin down more precisely the open features. We
then apply the set of newly adjusted parameterizations to several more
detailed observables such as neutron skin, isotope shifts, and super-heavy
elements. The techniques of least-squares fitting provide safe
estimates for the uncertainties of such extrapolations. The
systematic variation of forces allows to disentangle the various
influences on a given observable and to estimate the predictive value
of the SHF model. The results depend very much on the observable
under consideration.
\end{abstract}

\maketitle

\section{Introduction}

The production of exotic nuclei far off the valley of stability is
making steady progress at various laboratories around the world and is
providing an increasing amount of new experimental data on basic
nuclear properties, see, e.g., \cite{Naz03a,Mit04aR}. This is a great challenge
for nuclear structure theory at all levels of refinement, from the more
phenomenological microscopic-macroscopic methods (see,
e.g., \cite{Moe95aR}) through self-consistent mean-field (SCMF) methods
(see, e.g., \cite{Ben03aR}) or large shell-model calculations (see, e.g.,
\cite{Koo97aR}) up to several ab initio techniques employing a given
nucleon-nucleon interaction (see, e.g., \cite{Dea05a,Rot05a,For05a}).
Further development of all these methods is a highly topical task for
which a large network of activities has been launched recently
\cite{BerUNEDEF}. The present paper aims to contribute to the
development of SCMF.

There are several approaches to SCMF from which the most widely used
are the Skyrme-Hartree-Fock (SHF) model \cite{Ben03aR}, the Gogny
force \cite{Dec80a}, and the relativistic mean-field model (RMF)
\cite{Rei89aR,Rin96aR}. The genuine nucleon-nucleon interaction does
not allow an immediate mean-field treatment. Thus all SCMF models
employ effective interactions which are arranged to provide reliable
nuclear structure properties and low-energy excitations at the level
of a mean-field description. That approach has much in common with the
density functional theory widely used in the physics of electronic
systems \cite{Dre90aB}. The difference is, however, that electronic
correlations are well under control and that reliable electronic
energy-density functionals can be derived from well controlled ab
initio calculations.  The nuclear case is much more involved because
the nucleon as such is a composite particle and a nucleon-nucleon
interaction is already an approximate concept. Thus, nuclear many-body
theories have not yet reached sufficient descriptive power to serve as
direct input for deducing effective energy-density functionals for
SCMF. Although there are promising attempts for an ab initio
derivation \cite{Cao06a}, the general strategy for constructing energy
functionals for high-quality calculations is to deduce the formal
structure from principle considerations \cite{Neg72a} and to rely on a
phenomenological adjustment of the model parameters.  The present
paper aims at a critical and thorough survey of the SHF model and the
\PGRcomm{phenomenological adjustment} of its parameters. There is a
long history of SHF development and optimization (for recent reviews
see \cite{Ben03aR,Sto07aR}) and still need for further intense
consideration, as can be seen from the lively discussion in the
literature, recent re-fitting attempts \cite{Sam02a,Ber05a}, and the
huge combined effort in \cite{BerUNEDEF}. \PGRcomm{This study will
point to several incompatibilities with the present SHF ansatz when
trying to cover more than just the basic ground-state properties. We
are thus not yet at the stage to advertise one preferred parameterization.
The strategy is rather to supply a toolbox of forces produced under 
systematically varying conditions. 
}

After a brief review of the basic SHF functional and the strategy of
least-squares fitting, our considerations start with an inspection of
the input data for the adjustment of the model parameters.  The idea
is to select those ground-state observables (energies, radii, surface
thicknesses) which are expected to be well described by a pure
mean-field model. The limits to a mean-field description are set by
the ground-state correlations (GSC) stemming from low-lying collective
quadrupole modes, for recent surveys see \cite{Ben05a,Ben06a,Klu08a}.
Taking the general trends of GSC as worked out in \cite{Klu08a}, we
determine \PGRnew{a set of observables which have only small
correlation effects and which we use as basis for the fits of
mean-field models}. We then will investigate in detail how predictive
such fits can be. To that end we consider series of fits with
additional features added as constraint. These features are quantified
in terms of well known nuclear matter properties (NMP) like
incompressibility $K$, effective mass $m^*/m$, symmetry energy
$a_\mathrm{sym}$, and sum-rule enhancement factor $\kappa$
\cite{Ben03aR}. We study systematic variations of these NMP. Further
information from giant resonances (GR) is invoked to define the
optimal values for the NMP more precisely. Finally, we apply the set
of SHF parameterizations thus obtained to a study of the predictive
value for more detailed {properties} like, e.g., neutron skin, isotope
shifts of charge radii, and super-heavy elements.

\section{The Skyrme energy functional}
\label{sec:framework}

\newcommand{\bsigma}{\mbox{\boldmath$\sigma$}}
\newcommand{\btau}{\mbox{\boldmath$\tau$}}
\newcommand{\half}{{\textstyle\frac{1}{2}}}

The goal of a nuclear mean-field theory is to describe the
many-body system exclusively in terms of a set of single-particle
wavefunctions together  with the BCS occupation  amplitudes
$\left\{\varphi_\alpha,v_\alpha,\alpha=1,...,\Omega\right\}$.
 Thus the typical
mean-field state is a BCS state
\begin{equation}
  |\Phi\rangle
  =
  \prod_{\alpha>0}\big(
    u_\alpha + v_\alpha \hat{a}^+_\alpha \hat{a}^{+}_{\bar\alpha}
  \big)|0\rangle
\label{eq:BCState}
\end{equation}
where $|0\rangle$ is the vacuum state and
$u_\alpha=\sqrt{1-v_\alpha^2}$. The product
runs over all pairs of time-reversed partners $(\alpha,\bar\alpha)$ indicated
by $\alpha>0$.
The mean field equations are obtained by variation of the total energy
with respect to {single-particle
wave functions and pairing amplitudes}.  The expression for the total
energy is the key ingredient in the modeling. We will here employ the
Skyrme energy-density functional together with a pairing functional.
\PGRnew{Both functionals are} expressed in terms of a few local
densities and currents: local density $\rho$, kinetic-energy density
$\tau$, spin-orbit density ${\bf J}$, current ${\bf j}$, spin density
${\bsigma}$, {kinetic spin density ${\btau}$, and pair current
$\xi$}.  All appear twice, for protons and for neutrons, e.g.,
$\rho_p$ and $\rho_n$. In detail they read 
\begin{subequations}
\label{eq:basdens}
\begin{eqnarray}
  \rho_q({\bf r})
  &=&
  \sum_{\alpha\in q}v_\alpha^2\,
  \big|\varphi_\alpha({\bf r})\big|^2
  \;\,\;
  q\in\{{\rm p,n}\}
  \quad,
\\
  \tau_q({\bf r})
  &=&
  \sum_{\alpha\in q}v_\alpha^2\,
  \big|\nabla\varphi_\alpha({\bf r})\big|^2,
\\
  {\bf J}_q({\bf r})
  &=&
  -\mathrm{i}
  \sum_{\alpha\in q}v_\alpha^2\,
  \varphi_{\alpha}^+({\bf r})\nabla\!\times\!\hat{\mathbf\sigma}
  \varphi_{\alpha}^{\mbox{}}({\bf r})
  \quad,
\\
  {\bf j}_q({\bf r})
  &=&
  -{\textstyle\frac{\mathrm{i}}{2}}
  \sum_{\alpha\in q}v_\alpha^2\,
  \left(
   \varphi_{\alpha}^+({\bf r}) \nabla\varphi_{\alpha}^{\mbox{}}({\bf r})
   -
   \mathrm{c.c.}
  \right)
  \quad,
\\
  {\bsigma}({\bf r})
  &=&
  \sum_{\alpha\in q}v_\alpha^2\,
  \varphi_{\alpha}^+({\bf r})
  \hat{\bsigma}\varphi_{\alpha}^{\mbox{}}({\bf r})
  \quad,
\\
  {\btau}({\bf r})
  &=&
  \sum_{\alpha\in q}v_\alpha^2
  \sum_{i\in\{xyz\}}
  \nabla_i\varphi_{\alpha}^+({\bf r})
  \hat{\bsigma}\nabla_i\varphi_{\alpha}^{\mbox{}}({\bf r})
  \quad,
\\
  \xi_q({\bf r})
  &=&
  2\sum_{\alpha\in q}^{\alpha>0} u_\alpha v_\alpha
  \big|\varphi_\alpha({\bf r})\big|^2
  \quad.
\end{eqnarray}
It is often useful to recouple to sum and difference, e.g.,
\begin{equation}
  \rho
  =
  \rho_{\rm p}+\rho_{\rm n}
  \quad,\quad
  \tilde\rho
  =
  \rho_{\rm p}-\rho_{\rm n}
  \quad,
\end{equation}
\end{subequations}
and similarly for all other densities and currents.  The sum plays a
role in the isoscalar terms of the energy functional and we will call
it the isoscalar density $\rho$. In a similar manner, the difference
plays the role of an isovector density $\tilde\rho$.

Our starting point is then the most general Skyrme energy functional
\begin{subequations}
\label{eq:enfun}
\begin{eqnarray}
  E
  &=&
  \int d^3r\,\left\{
    {\mathcal E}_{\rm kin}
    +{\mathcal E}_{\rm Skyrme}
  \right\}
\nonumber\\
  &&
  +
  E_{\rm Coulomb}
  +
  E_{\rm pair}
  +
  E_{\rm cm}
  \quad,
\\
  {\mathcal E}_{\rm kin}
  &=& 
  \frac{\hbar^2}{2m_\mathrm{p}}\tau_\mathrm{p}
    +
    \frac{\hbar^2}{2m_\mathrm{n}}\tau_\mathrm{n}
  \quad, 
\label{eq:ekindens}
\\
  {\mathcal E}_{\rm Skyrme}
  &=&
  \frac{B_0+B_3\rho^\alpha}{2}\rho^{2} 
  - 
  \frac{B_0^{\prime}+B'_3\rho^\alpha}{2}\tilde\rho^2
\nonumber\\
  &&    
  +B_1 \big(\rho\tau-{\bf j }^2\big)
  -B_1^{\prime}\big(\tilde\rho\tilde\tau-\tilde{\bf j}^2\big)
\nonumber\\
  &&
  -\frac{B_2}{2}\rho\Delta \rho
  +\frac{B_2^{\prime}}{2}\tilde\rho\Delta\tilde\rho
\nonumber\\
  &&
  -\half B_4\left(
    \rho\nabla\!\cdot\!{\bf J} 
    +
    {\bsigma}\!\cdot\!(\nabla\!\times\!{\bf j})
   \right)
\nonumber\\
  &&
  -\half(B_4\!+\!b_4^{\prime})\left(
    \tilde\rho\nabla\!\cdot\!\tilde{\bf J} 
    +
    \tilde{\bsigma}\!\cdot\!(\nabla\!\times\!\tilde{\bf j})
   \right)
\nonumber\\
  &&
  +
   \frac{C_1}{2}\left({\bf J}^2-\bsigma\!\cdot\!\btau\right)
   - 
   \frac{C_1^\prime}{2}
   \left(\tilde{\bf J}^2-\tilde{\bsigma}\!\cdot\!\tilde{\btau}\right)
  \quad, 
\label{eq:enfunsk}
\\
  {E}_{\rm Coulomb}
  &=& 
  e^2 \frac{1}{2} \int d^3r\,d^3r'
        \frac{\rho_{\rm p}({\bf r })\rho_{\rm p}({\bf r}')}
             {|{\bf r }-{\bf r }'|}
\nonumber\\
  &&
      - \frac{3}{4}e^2\left( \frac{3}{\pi} \right)^{1/3}
        \int d^3r\big[\rho_{\rm p}\big]^{4/3}
  \quad,
\label{eq:coulfun}\\
  E_{\rm pair}
  &=&
  \frac{1}{4} \sum_q v_{0,q} \int d^3r \xi^2_q
  \big[1 -\frac{\rho}{\rho_\mathrm{pair}}\big]
  \quad,
\label{eq:ep2}\\
  E_{\rm cm}^{\mbox{}}
  &=&
  -\frac{1}{2mA}\langle\big(\hat{P}_\mathrm{cm}\big)^2\rangle
  \quad,\quad
  \hat{P}_\mathrm{cm}
  =
  \sum_i\hat{p}_i
  \ .
\label{eq:cmfull}
\end{eqnarray}
Accounting for the slight difference between the proton and the 
neutron mass ($\hbar^2/{2m_p}= 20.749821, \hbar^2/{2m_n} =  20.721260$) 
becomes important for exotic and heavy nuclei. 
The $B$ ($B'$) parameters determine the strength of the isoscalar
(isovector) forces.
The principle Skyrme functional ${\mathcal E}_{\rm Skyrme}$ contains
just the minimum of time-odd currents and densities which is required
for Galilean invariance \cite{Eng75a}, namely the
combinations $\rho\tau-{\bf j }^2$ and
$\rho\nabla\!\cdot\!{\bf J} 
 +{\bsigma}\!\cdot\!(\nabla\!\times\!{\bf j})$. 
Further conceivable time-odd couplings play a role only for odd nuclei 
and magnetic (unnatural parity) modes \cite{Ben02c}. 
\PGRcomm{
These additional time-odd terms are often derived 
starting with a density-dependent zero-range force.
We take here the point of view of density-functional theory
and start from an energy functional, choosing the minimalistic
approach in the time odd channel. That modeling is similar
to the strategy which  is also pursued in the relativistic
mean-field model \cite{Ser86aR,Rei89aR,Rin96aR}.
}
The parameters $B_i, B'_i$ are more convenient for the functional form
(\ref{eq:enfun}). They are uniquely related to the widely used
standard Skyrme parameters $t_i, x_i$ through
\label{eq:bdef}
\begin{alignat}{2}
B_0  & = \tfrac{3}{4}  t_0
         \quad , & \quad
 \nonumber \\
B'_0 & = \tfrac{1}{2}  t_0 (\half + x_0)  \quad , \nonumber \\
B_1  & = \tfrac{3}{16} t_1 + \tfrac{5}{16} t_2 + \tfrac{1}{4} t_2 x_2
         \quad , & \quad
 \nonumber \\
B'_1 & = \tfrac{1}{8} \Big[ t_1 (\half+x_1)-t_2 (\half+x_2)
                      \Big] \quad , \nonumber \\
B_2  & =   \tfrac{9}{32} t_1 
         - \tfrac{5}{32} t_2
         - \tfrac{1}{8}  t_2 x_2
         \quad , & \quad
 \nonumber \\
B'_2 & = \tfrac{1}{16} \Big[ 3t_1 (\half+x_1)+t_2 (\half+x_2)
                      \Big] 
         \quad , \nonumber \\
B_3  & = \tfrac{1}{8} t_3  
         \quad , & \quad
 \nonumber \\
B'_3 & = \tfrac{1}{12} t_3 (\half+x_3) 
         \nonumber \\
B_4  & = \half t_4 - \half b_4'
         \quad , 
\nonumber\\
C_1  & = \eta_\mathrm{tls}
         \tfrac{1}{8} \Big[   t_1 \big( \half - x_1 \big)
                            - t_2 \big( \half + x_2 \big)
                      \Big]
         \quad , & \quad
 \nonumber \\
C'_1 & = - \eta_\mathrm{tls}\tfrac{1}{16} ( t_1 - t_2 )
         \quad . \nonumber \\
\end{alignat}
\end{subequations}
Note the two non-standard entries. There is an extra parameter $b'_4$
for tuning the isovector dependency of the spin-orbit interaction.
A zero-range two-body spin-orbit
interaction leads to the fixed relation $b'_4=t_4/2$ while a
spin-orbit structure as suggested by the RMF is associated with
$b'_4=0$ \cite{Rei95a}. We use it here as free parameter.  The tensor
spin-orbit term associated with the terms $\propto C_1^{\mbox{}},C'_1$ is
omitted in many parameterizations. Here we introduce a new parameter
$\eta_\mathrm{tls}$ as a switch factor {where
$\eta_\mathrm{tls}=1$ includes the full tensor spin-orbit and
$\eta_\mathrm{tls}=0$ selects the widely used option to ignore the tensor
spin-orbit term}.
The Coulomb functional (\ref{eq:coulfun}) depends only on the charge
density and stays outside this distinction. Its second term approximates
exchange in the Slater approximation \cite{Sla51}. Note that we use
the proton density in place of the charge density. This is a widely
used, more or less standard, approximation.
The center-of-mass correction (\ref{eq:cmfull}) is an approximation to
the full c.m. projection \cite{Schm91a}. It is applied a posteriori and
its contribution to the mean-field equations is neglected. There are
other recipes for the c.m. correction not considered here, see
\cite{Ben03aR,Sto07aR}.
The pairing functional (\ref{eq:ep2}) involves the pair current $\xi$
and can be derived from a density-dependent zero-range force. The
parameter $\rho_\mathrm{pair}$ determines the weight of density
dependence. The limit $\rho_\mathrm{pair}\longrightarrow\infty$ recovers
the pure $\delta$ interaction (DI) which is also called volume
pairing.  The general case is the density-dependent $\delta$
interaction (DDDI).  A typical value near matter equilibrium density
$\rho_\mathrm{pair}=0.16\,\mathrm{fm}^{-3}$ concentrates pairing to the
surface. {Thus it is often denoted as surface pairing}. We will
consider $\rho_\mathrm{pair}$ as a free parameter and it will lead to an
intermediate stage between volume and surface pairing.

The results of the BCS calculations
depend on the space of single-nucleon states taken into account,
called here pairing phase space. In fact, the cut-off is part
of the pairing description. It is provided by the phase-space weights
$w_\alpha$ in the above pairing functionals.
We use here a soft cutoff profile such as,
$
  w_\alpha
  =
  \left[1+
    \exp{\left((\varepsilon_\alpha-(\epsilon_F+\epsilon_{\rm cut}))
            /\Delta\epsilon\right)}
  \right]^{-1}
$
where typically $\epsilon_{\rm cut}=5\,{\rm MeV}$ and
$\Delta\epsilon=\epsilon_{\rm cut}/10$ \cite{Bon85a,Kri90a}.  This
works very well for all stable and moderately exotic nuclei.  For
better extrapolation ability away from the valley of stability, the fixed
margin $\epsilon_{\rm cut}$ may be modified to use a band of fixed
particle number $\propto N^{2/3}$ instead of a fixed energy band
\cite{Ben00c}.

\PGRcomm{ 
When checking the performance of the fits for various observables (see
section \ref{sec:observ}), we encounter also deformed configurations.
These are computed with a code allowing for axially symmetric and
reflection asymmetric configurations. The reflection asymmetry becomes
important for fission barriers in actinides.
For well deformed nuclei, the dominant part of the correlation energy
comes from angular-momentum projection. We have accounted for that at
the level of the Gaussian-overlap approximation (GOA) in a form which
provides correctly a smooth transition to spherical shapes where the
correction vanishes \cite{Rei78b,Rei87aR,Rei99b,Hag02a}. This amounts to
subtract 
%
\PKcomm{
$E_\mathrm{ZPE,rot}=
g(\langle\hat{J}^2\rangle/4)\langle\hat{J}^2\rangle/(2\Theta_\mathrm{rot})
$
}
where the interpolating function is
%
\PKcomm{
$
g(x)=x\partial_x\log\big(\int_0^1da\exp{(x(a^2-1))}\big)
$
}.
That correction is included in all following results dealing with
deformed configurations.
}

\section{Fitting strategy}

\subsection{Global quality measure $\chi^2$ and minimization}

The free parameters of the SHF ansatz are going to be determined by a
least squares fit. To that end, we build a global quality measure by
summing the squared deviations from the data as
\begin{equation}
  \chi^2
  =
  \sum_{\mathcal{O},\mathrm{nucl}}
  \chi_{\mathcal{O},\mathrm{nucl}}^2
  \;,\;
  \chi_{\mathcal{O},\mathrm{nucl}}
  =
  \frac{\mathcal{O}^\mathrm{(th)}_\mathrm{nucl}
        -
        \mathcal{O}^\mathrm{(exp)}_\mathrm{nucl}}
       {\Delta\mathcal{O}_\mathrm{nucl}}
  \;,
\label{eq:chi2}
\end{equation}
where $\mathcal{O}$ stands for one of the selected observables,
``nucl'' for a nucleus, the upper index ``th'' for a calculated value,
and ``exp'' for the experimental date. The denominator
$\Delta\mathcal{O}$ stands for the \PGRnew{{\em adopted}} error of that
observable. It renders each contribution dimensionless and regulates
the relative weights of the various terms. 
The experimental uncertainty is of
little help here {because the experimental precision of these
basic bulk observables is much better than what we can expect from
the mean-field description, particularly for the binding energy}. 
The limiting factor comes from theoretical
uncertainties, \PGRnew{i.e.}, the quality we can expect from a mean-field
description. We will briefly describe the observables in section
\ref{sec:obs} and the selection of nuclei together with the adopted
errors in section \ref{sec:adopterr}.

The total quality measure is a function of all model parameters, i.e.
$\chi^2=\chi^2(p_1,...,p_N)$. We search for those parameters which
minimize $\chi^2$. As standard technique we employ the $\chi^2$
minimization technique from Bevington \cite{Bev69aB,Fri86a},
complemented occasionally by Monte-Carlo sampling to enhance the
chances for ending up in the global minimum. 

The aim of this publication is to explore systematically the
various influences from key features as, e.g., incompressibility or
symmetry energy. We thus add constraints on such key features. That is
done most simply by adding the wanted features as additional
observables with very small adopted errors $\Delta\mathcal{O}$ to the
$\chi^2$.

The rules of $\chi^2$ fitting also provide information to estimate the
statistical errors for extrapolations to other observables.  Let us
consider some observable 
$A$. Its expectation value is a
function of the model parameters, i.e. $A=A(p_1,...,p_N)$. The
extrapolation  error then  becomes
\begin{equation}
  \Delta A
  =
  \sqrt{
    \sum_{i,j}
    \frac{\partial A}{\partial p_i}
      \left(\mathcal{C}^{-1}\right)_{ij}
    \frac{\partial A}{\partial p_j}
  }
  \quad,\quad
  \mathcal{C}_{kl}
  =
  \frac{\partial^2\chi}{\partial p_k\partial p_l}
  \quad.
\label{eq:extraperror}
\end{equation}
That is the allowed variation of the observable $A$ within the
ellipsoid of $\chi^2-\chi^2_\mathrm{min}\leq 1$, i.e. for all $\chi^2$
which stay at most one unit above the minimum. We will exploit that
feature to compute extrapolation errors for all observables not
included in the fit data.

\subsection{Selection of fit observables}
\label{sec:obs}

The total binding energy $E_\mathrm{B}$ of a nucleus is the most
immediate observable in self-consistent mean-field models. It is
naturally provided by the numerical solution of the mean-field
equations.
The next important observables are related to the spatial extension of
the nucleus. It can be accessed experimentally through elastic
electron scattering which yields the nuclear charge form factor
$F_\mathrm{ch}(q)$ and by Fourier transformation the nuclear charge
density $\rho_\mathrm{ch}(r)$ \cite{Fri82a}.  SHF calculations yield
first the proton and neutron densities which need to be folded with
the intrinsic electro-magnetic structure of the nucleons to obtain the
charge density and charge form factor \cite{Fri75a}.  Mean-field models
provide reliable information about the form factor at low momentum $q$
\cite{Rei92b,Rei94aR}. \PGRnew{This} region of the form factor at low momentum
can be described by three parameters \cite{Fri82a}, the r.m.s. radius
\begin{equation}
r^2_{\rm rms} 
= -\frac{3}{F_{\rm ch}(0)} 
   \frac{d^2}{dq^2} F_{\rm ch} (q) \bigg|_{q=0} 
  ,
\end{equation}
the (first) diffraction radius
\begin{equation}
R 
= \frac{4.493}{q_0^{(1)}}
,
\label{eq:Rdif}
\end{equation}
which is determined from the first zero of the
form factor \mbox{$F_{\rm ch} (q_0^{(1)}) = 0$},
and the surface thickness 
\begin{equation}
\sigma^2 
= \frac{2}{q_m} \log{\left(\frac{F_{\rm box}(q_m)}{F_{\rm ch}(q_m)}\right)} 
\ , \quad q_m = 5.6/R  
.
\label{eq:surf}
\end{equation}
Therein $F_{\rm box}(q)$ corresponds to the form factor of a homogeneous
box of radius $R$, i.e. $F_{\rm box}(q)=3j_1(qR)/(qR)$.
The diffraction radius parameterizes the overall diffraction pattern
which resembles that of a box of radius $R$. It is the box-equivalent
radius. For more details find a recent summary in \cite{Ben03aR}.

The most prominent observable for \mbox{$T=1$} pairing correlations on
the mean-field level is the odd-even staggering of nuclear masses from
which {an approximation to the} pairing gap can be extracted,
e.g., using a five-point formula yields for the neutron gap
\begin{subequations}
\label{eq:gapdef}
\begin{eqnarray}
  \Delta_n^{(5)}
  &=&
  -
  \frac{1}{8}E(Z,N+2)
  +
  \frac{1}{2}E(Z,N+1)
  -
  \frac{3}{4}E(Z,N)
\nonumber\\
  &&
  +
  \frac{1}{2}E(Z,N-1)
  - 
  \frac{1}{8}E(Z,N-2)
\end{eqnarray}
and similarly for the proton gap $\Delta_p^{(5)}$.  
The computation of odd
nuclei using mean-field models, however, is very involved. {It requires
deformation and a long search for the optimum blocked configuration
which increases numerical expense by two orders of magnitude. Thus it
is} not well suited for systematic surveys.  One can compute an
average pairing gap from the state-dependent gap $\Delta_\alpha$ using a
weight which is sensitive to the region about the Fermi surface
\cite{Sau81a,Ben00c,Dug02b}
%
\begin{equation}
 \bar{\Delta}_q
 = 
  \frac{\sum_{\alpha\in q} u_\alpha v_\alpha \Delta_\alpha}
       {\sum_{\alpha\in q} u_\alpha v_\alpha}
\qquad (\alpha>0)
\end{equation}
\end{subequations}
These spectral gaps $\bar{\Delta}_q$ are found to be fairly well 
related to the five-point
gaps $\Delta_q^{(5)}$ in mid-shell regions \cite{Ben00c,Dug02b}.  For
magic nuclei and next to \PGRnew{them}, these two quantities develop very
differently \cite{Ben02b}. There remains some influence from nuclear
shape fluctuations on  $\Delta_q^{(5)}$ \cite{Dob01a} which gives an
uncertainty of about 10--20\% for the comparison of the two different
definitions (\ref{eq:gapdef}) for a gap. 

We also need some information which is specific to the spin-orbit
terms in SHF. The obvious quantity for the purpose are the l*s
splittings of single-particle energies.  Experimental information on
single-particle energies of even-even nuclei is drawn from the
single-nucleon removal energies or from the low-lying excited
states of the adjacent odd-$A$ nuclei.  That identification requires
that the polarization effects induced by the extra nucleon (or hole)
are small.  The magnitude of these effects have been investigated for
doubly-magic nuclei by \cite{Rut98a} within the RMF and \cite{Ber80a}
with SHF using the linear response theory.  It is found that energy
differences amongst particle states and hole states separately are
robust. Thus spin-orbit splittings are a robust signal as long as they
do not cross the shell gap. Furthermore, the involved single-particle
energies should not be too far away from the Fermi energy to keep
perturbation from higher configurations (additional $1ph$ couplings)
low. Another limitation has to be considered for states in the
particle spectrum because the state should stay safely off the
particle continuum. Thus we decide {to include only}
spin-orbit splittings of hole states close to the Fermi energy in
doubly-magic nuclei. The chosen data are listed in tables
\ref{tab:fitdata1} and \ref{tab:fitdata2}.

\subsection{Correlation effects and adopted errors}
\label{sec:adopterr}

\begin{figure*}
\centerline{\includegraphics[width=1.0\textwidth]{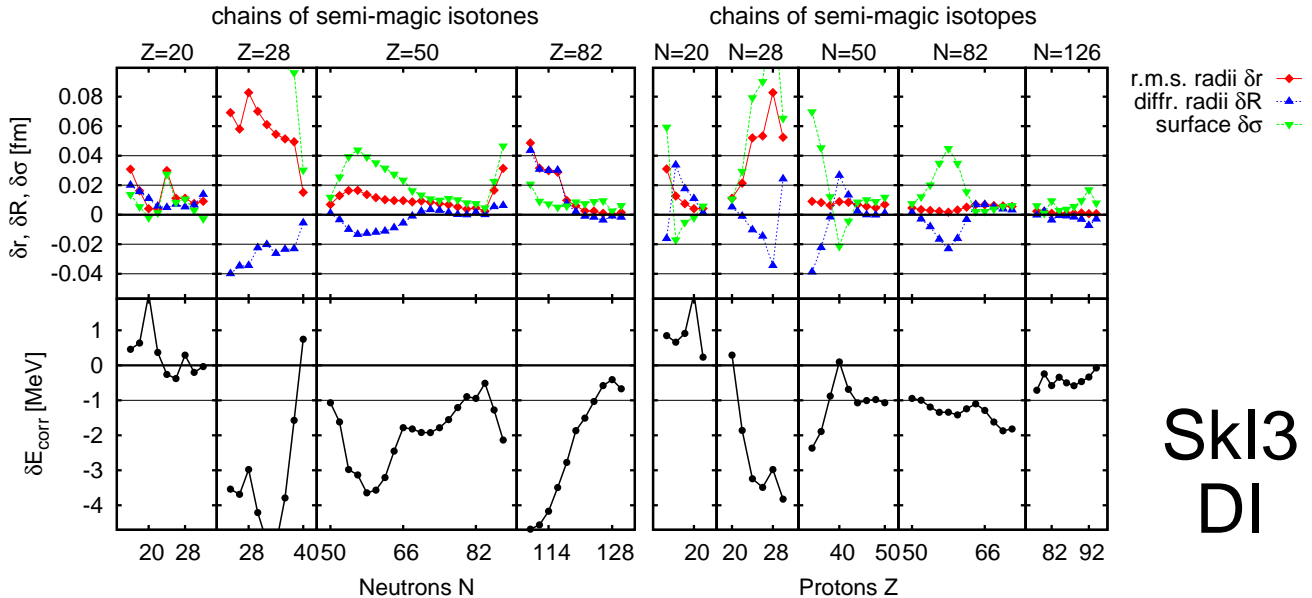}}
\caption{\label{fig:corr_all_ski3_DI_PNC_bw}
  Correlation effects on semi-magic isotopic (left) and isotonic
  (right) chains computed with SkI3 and DI pairing.
  The isotopic chains are Z=20, Z=28, Z=50, and Z=82, and the
  isotonic chains N=20, N=28, N=50, N=82, and N=126. 
  Upper panels: correlation shifts of charge r.m.s. radii
  \PGRnew{($\delta r$)}, diffraction radii,
  \PGRnew{($\delta R$)}, and surface thicknesses
  \PGRnew{($\delta\sigma$)}.
  Lower panels: correlation energies
  \PGRnew{($\delta E_\mathrm{corr}$)}.
  Horizontal dotted lines indicate intended error limits.
}
\end{figure*}
The {goal} is to adjust an effective nuclear energy-density
functional. The observables and nuclei included in the adjustment
should be well adapted for a pure mean-field description. Thus we have
to scrutinize the most dominant effects going beyond mean field. As
pointed out in the introduction, the strong short-range correlations
as well as correlations from resonances and high modes follow a trend
smooth in nucleon numbers $Z$ and $N$ and can be assumed to be effectively
incorporated into the energy-density functional \cite{Rei94aR}. We have
to care, however, about correlation effects which vary strongly within
the chart of isotopes. These stem predominantly from the low-lying
quadrupole excitations, associated with large amplitude collective
motion like soft vibrations and rotations, for extensive recent
analysis see \cite{Ben05a,Ben06a,Klu08a}.
\PGRnew{We are concerned here with the influence of correlations on
bulk observables, the change in energy 
$\delta E_\mathrm{corr}=E_\mathrm{corr}-E_\mathrm{mf}$
(where $E_\mathrm{corr}$ is the energy from the correlated calculation
and $E_\mathrm{mf}$ the mean field result), and similarly the
changes $\delta r$ in charge r.m.s. radii, $\delta R$ in diffraction
radii,  and $\delta\sigma$ in surface thickness.
}
 
A good chance for small correlation effects exists for semi-magic
nuclei where either the proton or the neutron number corresponds to a
shell closure. These nuclei are generally spherical. 
Figure \ref{fig:corr_all_ski3_DI_PNC_bw} summarizes the correlation
effects on binding energies (lower panels) and radii as well as
surface thicknesses (upper panels). The results were obtained with
the parameterization SkI3 \cite{Rei95a}. Other parameterizations
yield very similar results \cite{Klu08a} such that {figure
\ref{fig:corr_all_ski3_DI_PNC_bw} is a typical result for the error}
distribution for any reasonable Skyrme force.  The correlation effects
generally shrink with increasing system size, which is expected 
{because low-lying excitation energies generally
decrease}. It is a bit surprising, however, that the correlation
energies in isotopic chains {can grow so large in the
mid-shell region}. The hitherto
often underestimated isotonic chains are visibly less perturbed.
Light nuclei generally show larger fluctuations,
sometimes even acquiring unphysically positive correlation energies.
The small positive values about 0.2--0.5 MeV are still within the
precision of our method and imply practically negligible correlation
energies. The only exception is $^{40}$Ca at (Z=N=20). The
unreasonably large value is due to the fact that the $2^+$ mode is not
really collective in that nucleus \cite{Klu08a}\PGRcomm{, but that we
compute correlations using the GOA which is not necessarily valid
for non-collective modes. This unphysical case, however, is}
anyway excluded from the fit data set due to the 
Wigner effect (for $N=Z$) \PGRnew{\cite{Sat97b}}.

The faint horizontal lines in figure \ref{fig:corr_all_ski3_DI_PNC_bw}
indicate the \PGRnew{error bands associated with the adopted errors},
$1$ MeV for $E_B$, 0.04 fm for $R_\mathrm{diff}$ and $\sigma$, and
0.02 fm for $r_\mathrm{rms}$.  Observables which stay below these
limits are included in the fit data.  Points which {are outside the
desired error bands, but not too far away,} are included with degraded
error weight which is done by multiplying the general adopted error
$\Delta\mathcal{O}$ for that nucleus with a certain factor depending
on how large the expected correlation could be. Moreover, all nuclei
with N=Z are excluded {already} because they carry a correlation
contribution from the Wigner energy \cite{Sat97b} and we do not yet
have reliable means to compute that correction.
The choice of fit data thus deduced is listed in tables
\ref{tab:fitdata1} and \ref{tab:fitdata2}. One finds therein also the
values for pairing gaps and spin-orbit splittings. We do not have any
reliable estimate for the errors in these observables. The pairing gap
is deduced from {the five-point difference of binding energies
\cite{Ben00c}}.
\PGRnew{The pairing gaps for nuclei whose energy has degraded error 
weight are modified by the same degradation factor for reasons of
consistency.
}
The overall error is tuned such that
the r.m.s. average $\chi$ from all pairing gaps is of the same order
as for the other observables.  A similar {strategy taking over
weight factors from binding energies} is applied for the spin-orbit
splittings. Altogether this amounts to an adopted error of 0.12 MeV
for the gaps and 10\% for the spin-orbit splittings, for details see
again tables \ref{tab:fitdata1} and \ref{tab:fitdata2}.

\PGRcomm{
Correlation energies are always negative (the few exceptions at low
$A$ are defects) and r.m.s. radii always grow, but the fits produce an
average with deviations lying on both sides. One would very much like
to leave a margin for correlations. A safe construction would
require a deeper knowledge of the systematics of correlation effects
and development of simple estimates of them. At the present stage, we
to fit to straightforward mean-field results and refrain from adding
empirical corrections. 
}

\subsection{More detailed observables}
\label{sec:observ}

There are many more nuclear properties which are interesting to look
at but which are not yet suited for inclusion in a systematic fit,
because the relation to experimental data is somewhat uncertain
(neutron radii, giant resonance frequencies, fission barriers) or too
cumbersome to compute (low lying collective states). We will look at
several such observables a posteriori:
the neutron skin in $^{208}$Pb,
the isotope shift of charge radii between $^{214}$Pb and  $^{208}$Pb,
the neutron level sequence near the Fermi surface in $^{132}$Sn,
extrapolation to super-heavy elements (SHE), and
giant resonances (GR). 
{Some of these observables, namely} the GR resonance peak
frequencies, are even used as additional selection criteria. We will
now explain their computation, while the other
observables are introduced in later sections.

The dominant excitation modes of the nucleus are the giant resonances
(GR). Their average peak position can be related to basic features.
Heavy nuclei show least spectral fragmentation and are best suited for
evaluating these averages. We will consider GR in $^{208}$Pb, in
particular the isoscalar giant monopole resonance (GMR), the isovector
giant dipole resonance (GDR), and the isoscalar giant quadrupole
resonance (GQR). The spectral strength distribution is computed by the
Random-Phase-Approximation (RPA) done self-consistently with the same
Skyrme interaction as was used for the ground state, for technical
details see \cite{Rei92a,Rei92b}. We use a large phase space on a
large spherical grid of 30 fm radius to achieve a sufficiently fine
discretization of the continuum \cite{Rei06c} for subsequent folding
with a Lorentzian of frequency dependent width
$\Gamma=
  \mbox{max}\left((\hbar\omega-8\,\mathrm{MeV})/3.5\,\mathrm{MeV},
  0.1\,\mathrm{MeV}\right) 
$.  
The linear $\omega$ dependence starts at neutron emission threshold
\PGRnew{with an} empirically adjusted slope. It simulates the escape
width and to some extent the collisional width. The strengths in
$^{208}$Pb all have one unique peak in the GR region frequency that can
easily be read off. Comparison with the experimental data will be done
with respect to that GR peak frequency. The experimental values are
$\hbar\omega_\mathrm{GMR}=13.7\,\mathrm{MeV}$,
$\hbar\omega_\mathrm{GDR}=13.6\,\mathrm{MeV}$, and
$\hbar\omega_\mathrm{GQR}=10.9\,\mathrm{MeV}$
\cite{Wou91aER,Vey70a,GDRdata}.

\subsection{Nuclear matter properties (NMP)}
\label{sec:nucmat}

Homogeneous nuclear matter describes the leading contributions (volume
terms) to nuclear properties. They are considered as useful
pseudo-observables characterizing the bulk properties of effective
interactions. There exist close relations between such bulk
properties and certain combinations of Skyrme parameters, but it is
\PGRnew{often more instructive}
to discuss a parameterization in terms of these 
\PGRnew{nuclear matter properties (NMP)}.
When performing systematic variations of forces, we will
consider a scan of dedicated values for selected 
NMP rather than simple Skyrme parameters.

The leading quantity is the energy per particle $E/A(\rho)$, often called
equation of state.  Its minimum $(E/A)_{\rm eq}$ at
saturation density $\rho_{\rm eq}$ defines the volume energy,
related to the equilibrium state of nuclear matter. Energy and
density in finite nuclei are modified by surface and shell effects but
always stay close to these guiding values.
\PGRnew{Furthermore, we discuss NMP related to excitations
(zero sound).}
The incompressibility at the saturation point is given by
%
\begin{equation}
  K_\infty
  = 
  9\,\rho^2 \, \frac{d^2}{d\rho^2} \,
       \frac{{E}}{A}\Big|_\mathrm{eq}
  \quad,
\label{eq:incompr}
\end{equation}
where $\rho=\rho_n+\rho_p$ is the total density.
It corresponds to the curvature at the minimum, and is related to
breathing modes like the giant monopole resonance \cite{Bla80aR}.
The symmetry energy coefficient is related to the isovector curvature
at the saturation point
\begin{equation}
a_{\rm sym}
= \frac{1}{2} \frac{d^2}{d(\rho_n-\rho_p)^2}
  \frac{{E}}{A} \bigg|_\mathrm{eq}
  .
\label{eq:asym}
\end{equation}
\PGRcomm{Finite nuclei are also sensitive to smaller densities
whose symmetry energy is characterized additionally by the slope
\begin{equation}
a'_{\rm sym}
= \frac{1}{2} \frac{d}{d(\rho_n+\rho_p)}\frac{d^2}{d(\rho_n-\rho_p)^2}
  \frac{{E}}{A} \bigg|_\mathrm{eq}
\quad.
\label{eq:asymp}
\end{equation}
}
The isoscalar effective mass $m^*$ is calculated as
\begin{equation}
  \frac{\hbar^2}{2m*}
  = \frac{\hbar^2}{2m}
    + 
    \frac{\partial}{\partial\tau} \frac{{E}}{A}\bigg|_\mathrm{eq}
 \quad.
\label{eq:effmas}
\end{equation}
The isovector effective mass is usually expressed as the enhancement
factor of the Thomas-Reiche-Kuhn sum rule \cite{Rin80aB}, which reads
\begin{equation}
  \kappa_{\rm TRK}
  = 
  \frac{2m}{\hbar^2}
  \frac{\partial}{\partial(\tau_n-\tau_p)} 
  \frac{{E}}{A}\bigg|_\mathrm{eq}
  \quad.
\label{eq:sumrule}
\end{equation}
Note the subtle difference between total derivatives in
eqs. (\ref{eq:incompr},\ref{eq:asym}) and partial derivatives in
eqs. (\ref{eq:effmas},\ref{eq:sumrule}). The latter take $E$ as
written in the Skyrme functional (\ref{eq:enfun}) considering all
densities and currents ($\rho$, $\tau$, ...) as independent variables
while the total derivative first expresses all quantities in terms of
the actual Fermi momentum $k_\mathrm{F}=k_\mathrm{F}(\rho)$, and thus
the density, before performing the derivative,
e.g., considering $\tau\longrightarrow\tau(\rho)$.

\section{Results and discussion}

\PGRcomm{

In this section, we will present the parameterizations obtained by
adjustment to the above selected set of data, optionally with an
additional constraining condition. The aim of the survey is to
explore the influence of varying conditions in a systematic manner. We
thus name the forces with the header ``SV'' and add different three
digits qualifiers to indicate the constraint for which a force was
adjusted. 
As outlined in section \ref{sec:framework}, the Skyrme energy
functional leaves some options to choose. The practical consequences
thereof will be discussed in section \ref{sec:options}. As a result,
we will employ the following standard choices: The tensor spin-orbit
is omitted, i.e. $\eta_\mathrm{ls}=0$, while isovector spin-orbit
coupling $b'_4$ and the cutoff density in the DDDI recipe
$\rho_\mathrm{pair}$ are allowed as free parameters in the fits.

}

Furthermore, we will often compare with a few typical
parameterizations from the existing literature: SkM$^*$ as a widely
used, \PGRcomm{meanwhile somewhat obsolete,} old standard \cite{Bar82a}.
It belongs to the second generation of Skyrme forces which 
for the first time delivered a high-precision description of nuclear ground
states \PGRcomm{(as compared to the first generation forces)}. It was
developed with an explicit study of surface energy and fission
barriers in semiclassical approximation. The set SLy6 and its cousins
have been developed with a bias to neutron rich nuclei and neutron
matter aiming at astrophysical applications \cite{Cha98a}.  SkI3 (and
SkI4) exploit the freedom of an isovector spin-orbit force to obtain
an improved description of isotopic shifts of r.m.s. radii in neutron
rich Pb isotopes which posed a severe problem to all conventional
Skyrme forces \cite{Rei95a}.
\PGRcomm{
Finally, we consider one representative of the series of forces
started in \cite{Sam02a}.  These forces are fitted predominantly to
binding energies, but employ an huge pool of nuclei including odd and
deformed ones.  For deformed nuclei a simple correction
for the angular momentum projection was added,  and an ad-hoc correction for the
Wigner energy in $N$=$Z$ nuclei \cite{Sat97b} was applied.  All these
forces are \PGRnew{are of comparable quality with respect to the
reproduction of the binding energies of finite nuclei.} 
They differ
in details of treatment or boundary conditions.  We choose here the
force BSk4 from \cite{Gor03a} because it has an effective mass of
$m^*/m=0.92$ which comes close to the typical values of our fits.

\subsection{Unconstrained fits}

\subsubsection{The fit to standard data: SV-min}

\begin{figure*}
\begin{center}
\includegraphics[width=15.0cm]{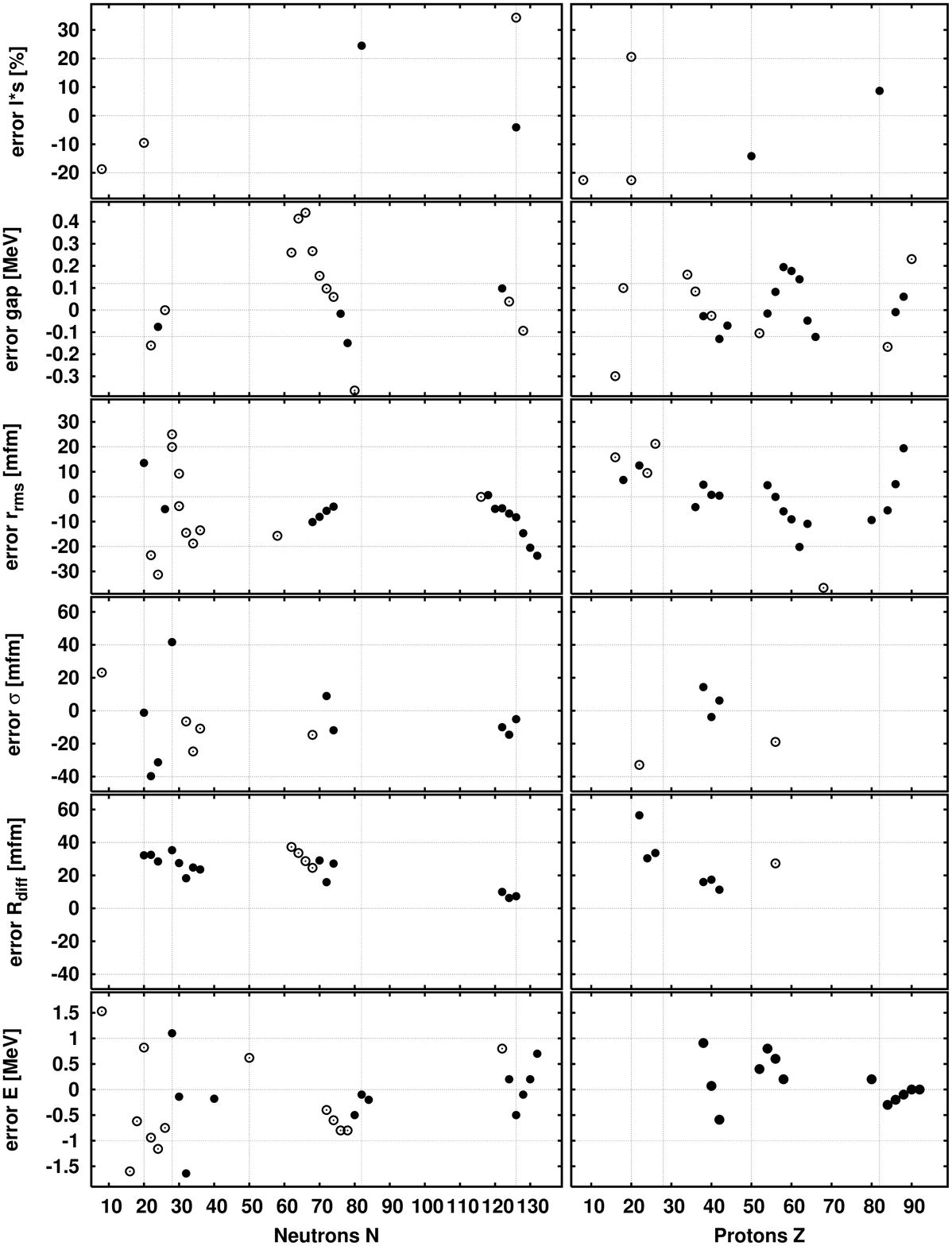}
\end{center}
\caption{\label{fig:SV-min-quality}
Deviation from experimental data for the nuclei and observables in the
fit sample as listed in tables \ref{tab:fitdata1} and
\ref{tab:fitdata2}. The results are drawn versus $N$ for the isotopic
chains (left) and versus $Z$ for the isotonic chains (middle).  The
values 0 (perfect matching) and the adopted error bands are indicated
by dotted horizontal lines.  Filled circles indicate data points 
fitted with full weight while open circles stand for data
points with reduced weight (weight factor $>1$ in tables
\ref{tab:fitdata1} and \ref{tab:fitdata2}).  Results are shown for the
parameterization SV-min resulting from a straightforward, unconstrained
fit to the data.
}
\end{figure*}
%
%
In a first round, we follow the most straightforward strategy and
adjust the model parameters to the data as selected in the previous
section and detailed in tables \ref{tab:fitdata1} and
\ref{tab:fitdata2}, without any additional constraint. The result is
called the parameterization ``SV-min'', for its detailed model
parameters see table \ref{tab:forces}.  Figure
\ref{fig:SV-min-quality} shows the deviation from the given data for
each observable and nucleus. All results stay fairly well within the
chosen error bands. That holds even for the points fitted with lower
weight (open circles).  In fact, the error bands are not fully
exhausted and the r.m.s. errors stay safely below the adopted errors.
The bands are generally filled on both sides of the zero line which
means that the fit averages nicely through the deviations. An
exception is here the diffraction radius (second panels from below)
where the deviation is always positive which indicates that this
observable is not easy to adjust within the given model. The
theoretical results are (within the allowed errors) systematically
larger than the experimental data.
Only few points exist for the spin-orbit splitting (uppermost panel)
and these do not fit as nicely as the other observables even with a
low demand such as a 20\% r.m.s. error. This indicates that single particle
structure is a very demanding observable. We will see that again for
other level sequences later on.

\begin{table}
\begin{tabular}{|l|rcll|}
\hline
 & \multicolumn{1}{|c}{$\chi^2$}  
 & \multicolumn{1}{c}{$\chi^2$/point}  
 & \multicolumn{2}{c|}{r.m.s. error} 
\\
\hline
binding energy $E_B$ & 12.07 &   0.17   &  0.62  MeV & (1.0) \\
diffr., radius $R$    & 11.18 &   0.40   &  0.029 fm & (0.04)\\
surface thick. $\sigma$ &  4.22  &  0.26 &  0.022 fm & (0.04)\\
r.m.s. radius $r$  &  15.86 & 0.32  &  0.014 fm & (0.02)\\
pairing gap $\Delta_p$ & 4.27 & 0.25  &  0.11  MeV& (0.12)\\
pairing gap $\Delta_n$ & 2.43 & 0.15  &  0.14  MeV& (0.12)\\
l$\cdot$s splitting & 3.18 &   0.45  & 0.25  \% & (20)\\
\hline
total  &  53.22  &  0.26   &&\\
\hline
\end{tabular}
\caption{\label{tab:SV-min-chi}
Global quality measures for various classes of observables
as achieved with the parameterization SV-min.
The second column shows the contribution from an observable
to $\chi^2$ while the third column expresses this as
$\chi^2$ per data point. The last column produces the
r.m.s. errors as such and the numbers in brackets indicate
the adopted error taken as weights for the fit, see eq. 
(\ref{eq:chi2}).
}
\end{table}
The global quality measures for the fit SV-min are shown in table
\ref{tab:SV-min-chi}. The r.m.s. errors (last column) show again what
we have seen in figure \ref{fig:SV-min-quality}, namely that
energies and form parameters (radii, surface thickness) perform very
well, better than the adopted errors as estimated from expected
correlation effects. It seems that a part of the correlations can be
accounted for by the model parameters. Pairing gaps are just at the
wanted limits and the l$\cdot$s splittings are somewhat at the edge.
The good overall performance yields a very low total $\chi^2$
as can be seen from the very small  $\chi^2$ per data point of
about 1/4. Typical fits aim at a  value of one.
The strict rules of  $\chi^2$-fitting would allow to reduce the
adopted errors until a $\chi^2$ per data point of one is reached.
\PGRnew{We are not pursuing that strategy here because our
adopted errors are determined by the}
expected reliability of mean field models from estimating correlation
effects. 
\PGRnew{The fact that $\chi^2$ per data point comes out much lower than
one indicates} the enormous versatility of
the Skyrme energy functional to describe global ground-state
properties.

\begin{figure}
\begin{center}
\includegraphics[width=7.8cm]{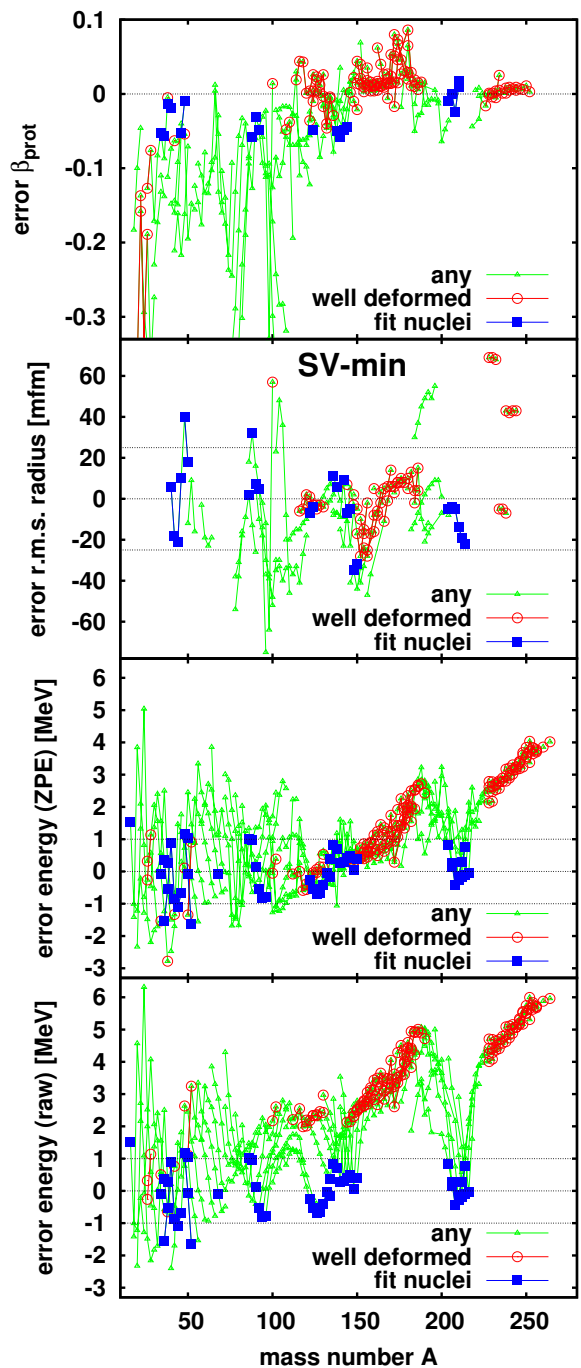}
\end{center}
\caption{\label{fig:SV-min-energies}
Errors in binding energies (lower two panels), r.m.s. radii (second
from above), and proton deformation 
throughout all nuclei for which experimental data were
available. The lowest panel uses the pure (deformed) mean-field value
for the binding energies while angular-momentum projection is
accounted for in the middle panel.
The experimental deformations are deduced from electro-magnetic
B(E2) values \cite{Ram01aR} and are compared with the
quadrupole variances (see section \ref{sec:observ}).
For energies and radii, 
the nuclei which were included in the fit are marked by filled
squares, well-deformed nuclei by open circles, and all others by
triangles.
}
\end{figure}
In order to check the interpolation and extrapolation properties, we
show in figure \ref{fig:SV-min-energies} the errors in binding
energies and r.m.s. radii throughout all known nuclei \PGRnew{including}
deformation. The energies are \PGRnew{displayed} in two ways. 
\PGRnew{The lowest panel shows the
energies
as they result straightforwardly from the mean-field calculations
(including c.m. correction)}. The fit
nuclei (filled squares) perform very well while large deviations can
develop for other nuclei. Note that the majority of deviations is
positive which indicates that additional binding through correlation
energies would correct in the desired direction. For well-deformed
nuclei, the dominant part of the correlation energy comes from
angular-momentum projection. We have accounted for that at the level
of the Gaussian-overlap approximation (GOA) \cite{Rei78b,Rei87aR,Klu08a}.
The results \PGRnew{with angular-momentum projection}
are shown in the middle panel of figure
\ref{fig:SV-min-energies}. The deformed nuclei up to Pb now perform
very well. The remaining discrepancies in the region $A<210$ are very
likely missing correlations. \PGRnew{This means} that the fit
interpolates nicely for all these nuclei provided soft nuclei are
computed with quadrupole ground-state correlations. However, for
actinides and super-heavy elements (SHE), the trend of the deviations is
too strong to be cured by remaining vibrational correlations.
The extrapolation to deformed super-heavy elements is plagued by a
growing trend to underbinding. We will take up that question later on.
The results for the r.m.s. radii look agreeable. The larger deviations
for some soft nuclei may still be cured by correlations. There are not
yet enough data to read off a trend for super-heavy elements.
The uppermost panel shows the difference between theoretical and
experimental proton deformations as derived from the B(E2) values.
They are given as dimensionless quadrupole moments associated with the
operator $\hat{\beta}=\hat{Q}_{20}\sqrt{\pi}/(\sqrt{5}Ar^2)$ where
$r$ is the r.m.s. radius and $\hat{Q}_{20}=r^2Y_{20}$ the spherical
quadrupole operator.
The theoretical values include the quadrupole variance, i.e.
$\beta^\mathrm{B(E2)}_\mathrm{theo}=
\sqrt{\langle\hat{\beta}^2\rangle}
=
\sqrt{\langle\hat{\beta}\rangle^2+\langle\Delta^2\hat{\beta}\rangle}
$.
It is to be remarked that this variance from the mean-field ground
state underestimates the true variance for soft vibrators and
transitional nuclei. This lets us expect huge deviations for the
majority of nuclei and this is indeed seen in the results.  However,
the picture looks better for rather rigid spherical nuclei with small
correlations (our fit nuclei) and well deformed nuclei.  
\PGRnew{It is a surprisingly}
nice agreement in view of the fact that sizeable contributions
to the variance from the true collective ground state are still missing.

Further properties of SV-min will discussed later in connection and
comparison with other parameterizations.

\subsubsection{A fit including a super-heavy nucleus}

\begin{figure}
\begin{center}
\includegraphics[width=7.8cm]{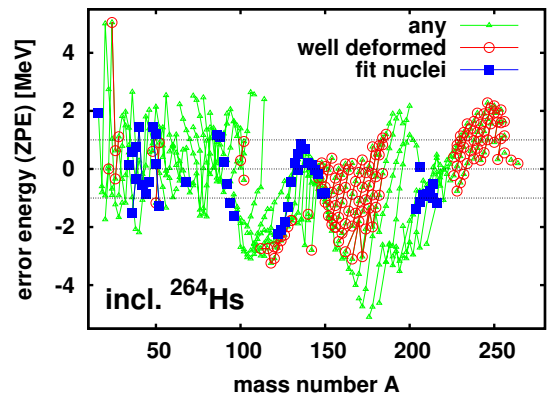}
\end{center}
\caption{\label{fig:SV-def-energies-2}
Errors in binding energies (including c.m. correction and
angular-momentum projection in GOA) throughout all nuclei for a
parameterization where the binding energy of $^{264}$Hs was 
\PGRnew{additionally included in the $\chi^2$.}
}
\end{figure}
We have seen above that the fitted force performs agreeably for
interpolations, but not so well for extrapolations. This suggests to
extend the range of interpolation by adding super-heavy nuclei to the
pool of data. Thus we have performed a fit where the binding energy
of $^{264}$Hs was added with high weight to the standard set of data.
That nucleus is well deformed. The angular-momentum projected energy
has been taken as reference value for the fit. The results for the
performance on energies are shown in figure
\ref{fig:SV-def-energies-2}. The error for $^{264}$Hs has successfully
been curbed down to almost zero.  But that is achieved at the price of
lowering all energies for medium-heavy and heavy nuclei. The
performance for the fit nuclei is visibly degraded and other nuclei
are turned to overbinding leaving no space for possible correlation
energies. We obviously encounter a deep-rooted problem with the given
Skyrme energy functional when considering the overall binding energy
of SHE.  The discussion of SHE will be continued in sections
\ref{sec:SHE} and \ref{sec:SHE2}.

\subsection{Fits with constraints on NMP}

\subsubsection{Variation of NMP}


\begin{table}
\begin{tabular}{l|rlll|rrr|r}
\hline
force & \multicolumn{1}{|c}{$K$}
      & \multicolumn{1}{c}{$m^*/m$} 
      & \multicolumn{1}{c}{$a_\mathrm{sym}$} 
      & \multicolumn{1}{c|}{$\kappa$} 
      & \multicolumn{1}{c}{$\rho_\mathrm{eq}$} 
      & \multicolumn{1}{c}{$E/A$} 
      & \multicolumn{1}{c}{$a'_\mathrm{sym}$}
      & \multicolumn{1}{|c}{$\chi^2$}
      \\
\hline
SV-min   &  222 &  0.95 & 30.7 & 0.08 &  0.1610 & -15.91 & 93 & 53.2\\
\multicolumn{1}{r|}{$\pm$}  
         &  8 & 0.15 & 1.4 &0.40 &  0.0013 & 0.06 & 89\\
\hline
\hline
SV-bas   &  234 &  0.9 & 30   & 0.4  &  0.1596 & -15.90 & 68 & 57.9\\
\hline
\hline
SV-K218 &  218 &  0.9 & 30   & 0.4  &  0.1615 & -15.90 & 72 & 57.0\\
\hline
SV-K226 &  226 &  0.9 & 30   & 0.4  &  0.1605 & -15.90 & 71 & 56.3\\
\hline
SV-K241 &  241 &  0.9 & 30   & 0.4  &  0.1588 & -15.91 & 65 & 61.3\\
\hline
\hline
SV-mas10 &  234 &  1.0 & 30   & 0.4  &  0.1594 & -15.91 & 59 & 60.8\\   
\hline
SV-mas08 &  234 &  0.8 & 30   & 0.4  &  0.1597 & -15.90 & 84 & 59.5\\
\hline
SV-mas07 &  234 &  0.7 & 30   & 0.4  &  0.1500 & -15.89 &109 & 68.0\\
\hline
\hline
SV-sym28 &  234 &  0.9 & 28   & 0.4  &  0.1595 & -15.86 & 15 & 63.4\\
\hline
SV-sym32 &  234 &  0.9 & 32   & 0.4  &  0.1595 & -15.94 &119 & 58.0\\
\hline
SV-sym34 &  234 &  0.9 & 34   & 0.4  &  0.1592 & -15.97 &169 & 61.6\\
\hline
\hline
SV-kap00 &  234 &  0.9 & 30   & 0.0  &  0.1598 & -15.90 & 82 & 52.6\\
\hline
SV-kap20 &  234 &  0.9 & 30   & 0.2  &  0.1597 & -15.90 & 74 & 56.5\\
\hline
SV-kap60 &  234 &  0.9 & 30   & 0.6  &  0.1595 & -15.91 & 61 & 60.5\\
\hline
\hline
SV-tls   &  234 &  0.9 & 30   & 0.4 &  0.1595 & -15.89 & 69 & 61.2\\
\hline
SkM$^*$  &  217 &  0.79 & 30  & 0.53 &  0.1602 & -15.75 &  95 & \\
SLy6     &  230 &  0.69 & 32  & 0.25 &  0.1590 & -15.92 & 100 &\\
SkI3     &  258 &  0.58 & 35  & 0.25 &  0.1577 & -15.96 & 212 &\\
BSk4     &  237 &  0.92 & 28  & 0.18 &  0.1575 & -15.77 &  27 &\\
\hline
\end{tabular}
\caption{\label{tab:nucmat}
NMP as defined in section \ref{sec:nucmat}
for the 
for the various SHF parameterizations used in  this paper
($K$  in MeV, $a_\mathrm{sym}$  in MeV, $E/A$   in MeV,
 $\rho_\mathrm{eq}$ in fm$^{-3}$,  $a'_\mathrm{sym}$
 in MeV\,fm$^3$, $m^*/m$ dimensionless, and $\kappa$ dimensionless).
The rightmost column lists the global quality measure $\chi^2$.
The parameterization SV-min results from an unconstrained
minimization of the total quality measure $\chi^2$ according to eq. 
(\ref{eq:chi2}) with the data and adopted errors from tables
\ref{tab:fitdata1} and \ref{tab:fitdata2}. The other parameterizations
{were obtained by} a fit constrained on four NMP. SV-bas
is the base point of the systematic variation with the constraints:
$K=234$ MeV,
$m^*/m=0.9$, $a_\mathrm{sym}=30$ MeV, and $\kappa=0.4$.
From that point, one property is varied, the incompressibility $K$ (via
power of density dependence $\alpha$) in SV-K, the effective mass 
in SV-mas, the symmetry energy in SV-sym, and the sum rule enhancement
in  SV-kap. Finally, SV-tls is constrained like SV-bas but employs
the full tensor spin-orbit terms.
Moreover, we append to the list the NMP for the four conventional
Skyrme forces used in several comparisons. 
}
\end{table}
The uppermost entry of table \ref{tab:nucmat} shows the NMP and the
final $\chi^2$ for SV-min together together with its extrapolation
uncertainties computed according to eq. (\ref{eq:extraperror}).  The
$\chi^2$ is very small in view of about 200 given data points.  The
NMP are more or less in commonly accepted ranges as used, e.g., in the
liquid-drop model \cite{Rei06a}. The ``ground-state'' properties,
equilibrium density $\rho_\mathrm{eq}$ and binding energy $E/A$, are
well fixed while all the other NMP related to excitations show
sizeable uncertainties. The fit leaves some freedom in these
respects. Moreover, we will see that the most prevailing nuclear
excitations, the giant resonances (GR), are not all so well tuned in
SV-min. This suggests to exploit the freedom left by the $\chi^2$ fits
for a fine tuning of GR. Moreover, it is interesting as such to
explore the large space of still allowed variations.  To that end, we
perform fits to the given set of data where additionally four NMP are
kept fixed: incompressibility $K$, effective mass $m^*/m$, symmetry
energy $a_\mathrm{sym}$, and sum rule enhancement factor $\kappa$ (see
section \ref{sec:nucmat} for its definition).  The first two are
isoscalar properties and the last two isovector.  One may wonder why
the slope of the symmetry energy, $a'_\mathrm{sym}$, is not included
in the variation. The reason is that $a'_\mathrm{sym}$ is tied up very
closely to $a_\mathrm{sym}$ by the fits such that only one of both
shows that freedom of choice. 

A four-dimensional landscape of variations of NMP is too bulky to
handle. We prefer to define one ``base point'' about which we perform
variation of one NMP at a time thus dealing with four sets of
variations. For the choice of the base point, we exploit the full
space of variations in the four NMP and use the freedom to accommodate
the GR properties as well as possible (see section \ref{sec:infoGR}).
After a longer search in that four-dimensional landscape, we find the
following choice a good compromise: $K=234$ MeV, $m^*/m=0.9$,
$a_\mathrm{sym}=30$ MeV, and $\kappa=0.4$.  \PGRnew{The corresponding}
parameterization is called ``SV-bas''. We then vary each one of the
four NMP while keeping the other three at the base value. This yields
four sets of variations with prefix SV-K for varied $K$, SV-mas for
varied $m^*/m$, SV-sym for varied $a_\mathrm{sym}$, and SV-kap for
varied $\kappa$. Table \ref{tab:nucmat} shows the NMP for these
parameterizations and the detailed parameters for the functional
(\ref{eq:enfun}) are provided in table \ref{tab:forces} in appendix
\ref{app:details}.
There is also one parameterization SV-tls with the
base values for NMP but now including tensor spin orbit, i.e
$\eta_\mathrm{tls}=1$. This serves to explore the effect of the tensor
spin-orbit term by comparison with SV-bas.

The overall quality measure $\chi^2$ is shown in the rightmost column
of table \ref{tab:nucmat}. The constraints, of course, degrade the
quality a bit. But we see that the variations yield $\chi^2$ which
stay within an acceptable range of about 10\% increase in $\chi^2$.
As we will see in section \ref{sec:infoGR}, SV-bas performs much
better than SV-min for GR in $^{208}$Pb. That counterweights the
small losses on the side of ground-state properties.

The NMP as given in table \ref{tab:nucmat} characterize ground states
and excitation properties for modes with natural parity. As typical
representatives for excitations with unnatural parity, we have also
checked spin modes in nuclear matter. These can be characterized by
the Landau parameters $g_0$ for pure spin excitations and $g'_0$ for a
spin-isospin mode \cite{Ost92aR}. Taking up the formula as given in
\cite{Ben02c}, we have computed these Landau parameters for the energy
functional (\ref{eq:enfun}). The only contribution comes from the
tensor spin-orbit term. Thus we have $g_0=0=g'_0$ for all forces with
$\eta_\mathrm{tls}=0$. Only SV-tls has a non-vanishing tensor
spin-orbit term. For that force we find $g_0=-0.73$ and
$g'_0=0.191$. These numbers stay safely above the critical value for
spin(-isospin) instability. Thus all forces introduced here are stable
in the spin channels. Note that we are here using the energy
functional with minimal time-odd terms, namely just those which are
required to achieve Galilean invariance.  That leaves the effective
interaction in all channels much more robust. Thus one should not be
puzzled by the rather large negative values for the spin-exchange
parameters $x_1$ and $x_2$ in table \ref{tab:forces}. These come from
expressing the functional in terms of the conventional Skyrme
parameters, mediated through eq. (\ref{eq:bdef}). Spin stability is
guaranteed in connection with the functional
(\ref{eq:enfun})
A different picture would evolve when taking the Skyrme force
literally as a zero-range force. This yields additional terms in the
spin channel which can easily render a parameterization unstable in
the spin channel. However, from an energy-density functional
viewpoint, we see no compelling reason to include those terms.

\subsubsection{Trends of the errors}

\begin{figure*}
\begin{center}
\includegraphics[width=17.0cm]{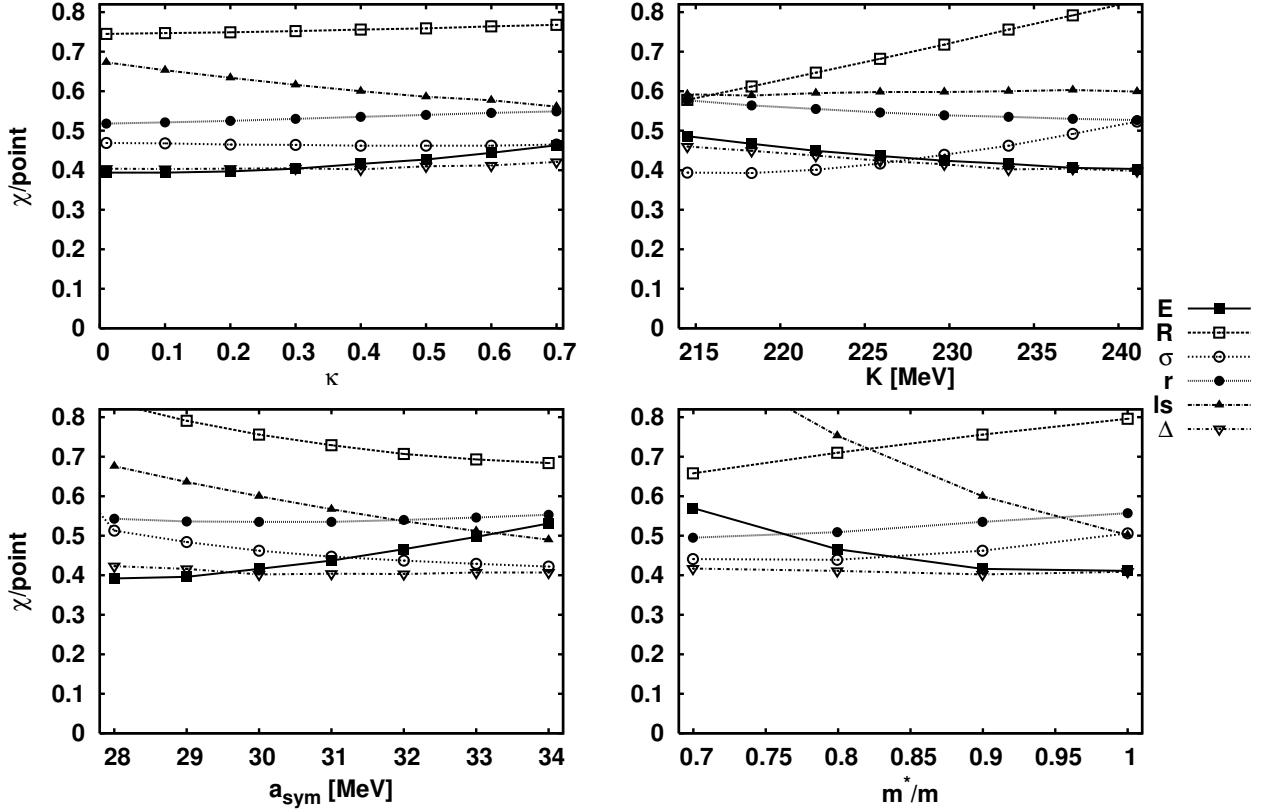}
\end{center}
\caption{\label{fig:TVabs_vary3_collect}
The distributions $\chi_\mathrm{obs}$ for each observable
as indicated 
for four cases of systematically varied bulk parameters.
Lower left: variation of symmetry energy $a_\mathrm{sym}$;
upper left: variation of sum rule enhancement $\kappa$;
lower right: variation of effective mass $m^*/m$;
upper right: variation of incompressibility.
}
\end{figure*}
The quality measure $\chi^2$ is composed of different observables
whose relative weight is determined by the given adopted errors.  It
is known from earlier studies that the choice of observables has an
influence on extrapolated NMP (see, e.g., \cite{Fri86a}). The sets with
systematically varied NMP now allow to visualize these trends.
}
Figure \ref{fig:TVabs_vary3_collect} shows the r.m.s. averaged
$\chi_\mathrm{obs}$ \PGRcomm{per data point} for each observable
separately. The figure has four panels to show the trends with respect
to the four NMP variations considered: incompressibility $K$,
effective mass $m^*/m$, symmetry energy $a_\mathrm{sym}$, and sum rule
enhancement $\kappa$. Variation of sum rule enhancement $\kappa$
changes very little in all observables. That feature is only loosely
determined by the data set, \PGRcomm{as already seen in table
\ref{tab:nucmat} from the rather large uncertainty for SV-min.}. We
will need further conditions to make a more definite choice. The other
three features lead all to sizeable trends, but often lead 
in different directions. For example, the lower right panel
shows that the r.m.s. radii would prefer low values of $m^*/m$ while
the energy prefers $m^*/m\approx 0.9$, \PGRcomm{other observables to
even higher $m^*/m$}. The final ``optimum'' for $m^*/m$ depends very
much on the choice of the relative weight of the different
observables. Even within the energy as observable, we could revert
the trend when giving light nuclei more weight by using relative
errors \cite{Fri86a} rather than absolute errors as done here.
Significantly different trends are seen also for variation of
$a_\mathrm{sym}$ and $K$. It is thus obvious that the relative weights
in the composition of the $\chi^2$ determine the final
\PGRcomm{extrapolated NMP}. Note that the actual changes in the
contributions to $\chi^2$ are small such that the total $\chi^2$
varies only very little when scrolling through the different
NMP. There is a broad choice of well performing
parameterizations. This explains why there are so many different SHF
parameterizations around which all provide a good description of
nuclear ground-state properties but vary in several of the key
features. In other words, the strategy of $\chi^2$ fitting leaves
several vaguely determined aspects. One needs to include further
observables which are more specific to the open features.

\subsection{Information from giant resonances}
\label{sec:infoGR}

\PGRcomm{

\subsubsection{GR in $^{208}$Pb and its relation to NMP}

\begin{figure*}
\begin{center}
\includegraphics[height=17.5cm,angle=-90]{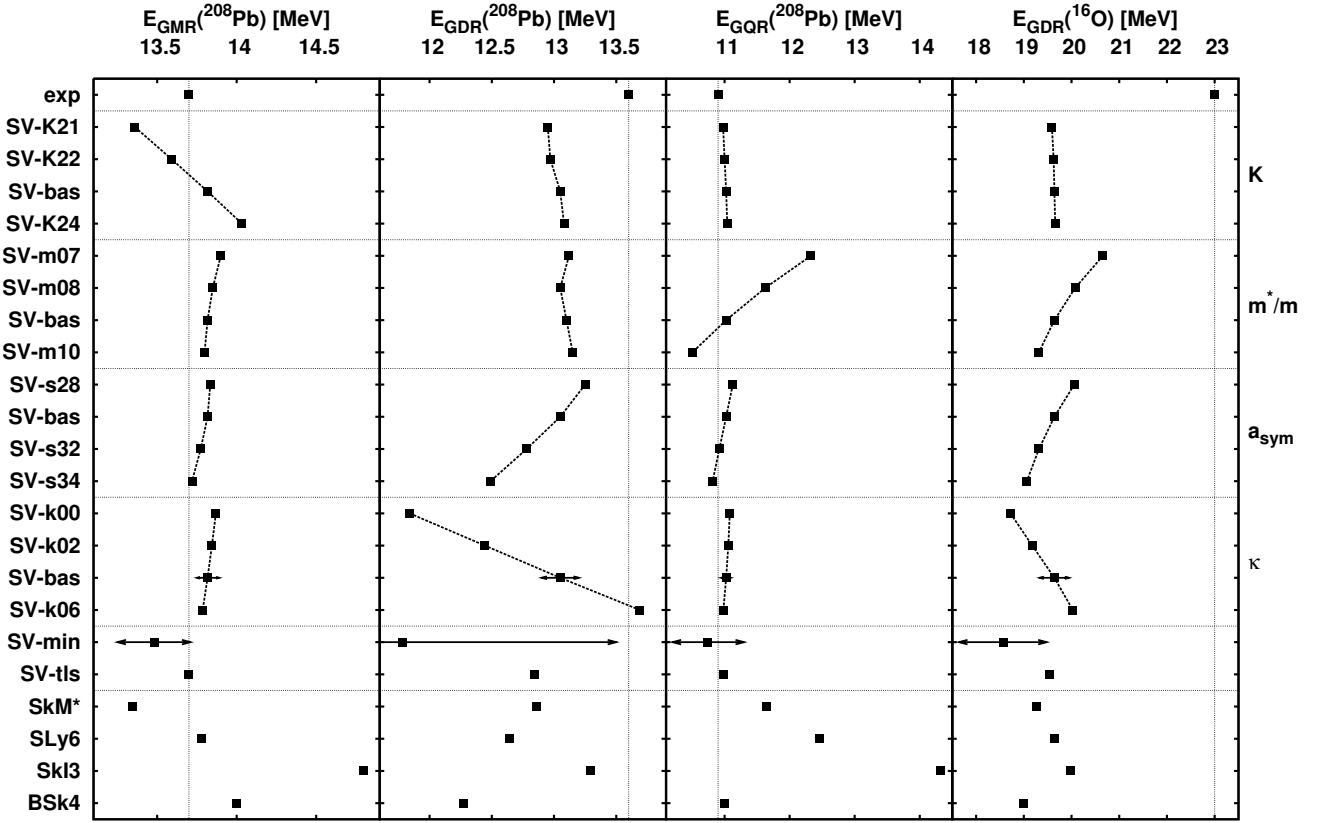}
\end{center}
\caption{\label{fig:vary-GR}
\PGRcomm{
Mean resonance frequencies for giant resonances in $^{208}$Pb 
(from left to right: GMR, GDR, and GQR) and for the GDR in
$^{16}$O computed with the various forces encompassing systematic  variation
of NMP properties
incompressibility $K$, 
effective mass $m^*/m$, symmetry energy $a_\mathrm{sym}$, and
sum rule enhancement $\kappa$ as indicated at the right side of the
panels.
Additionally, results from SV-min, SV-tls and traditional Skyrme
forces, as indicated, comprise the lower entries.
Extrapolation errors for the three observables
are shown as error bars for SV-bas and SV-min. Those for SV-bas
apply to all forces fitted with constrained NMP.
The experimental values are given in the uppermost entry and
drawn through as faint vertical lines.
}
}
\end{figure*}
Figure \ref{fig:vary-GR}
collects results for the peak frequencies of the GR in
$^{208}$Pb and of the GDR in $^{16}$O. We concentrate first on
discussing the GR in $^{208}$Pb.
The results for the straightforward fit SV-min are mixed.  The GQR
fits nicely, the GMR lies slightly to low, and the GDR is far off the
goal. The uncertainties (see error bars for SV-min) are sufficiently
large such that a good reproduction of GR in $^{208}$Pb seems within
the reach of allowed variations.

The relation between NMP and GR properties becomes apparent
from the various chains of the systematically varied forces.
} 
The situation is particularly simple for the isoscalar
excitations. There is a unique relation between an isoscalar GR and
isoscalar NMP: the GMR is sensitive exclusively to $K$ and the GQR to
$m^*/m$. Both these GR show a clean excitation spectrum with one
peak. Thus we use these two data points to fix the otherwise weakly
determined isoscalar NMP choosing $m^*/m=0.9$ to meet the GQR and
$K=234$ MeV to tune the GMR.
The case is much more involved for the isovector GDR which reacts to
two isovector NMP, to $a_\mathrm{sym}$ and to $\kappa$. Moreover, the
spectral distribution \PGRcomm{(not shown here)} tends to be strongly
fragmented, particularly for high $m^*/m$ and high $\kappa$. Lower
$m^*/m$ are excluded because we want to maintain the good adjustment
of the GQR. Thus we stay with a compromise for the GDR, choosing
$\kappa=0.4$ and $a_\mathrm{sym}=30$ MeV. This obviously does not
perfectly meet the experimental peak position for the GDR.  But the
example of the GDR in $^{16}$O discussed later on demonstrates that
the description of GDR by SHF is anyway not yet well under control.
The compromise here is to be understood as a preliminary setting, open
for the necessary further studies on the GDR.
\PGRcomm{Fixing these four settings in the fit yields the force SV-bas
as introduced in the previous section.  It serves as base point for
further variations of NMP. The extrapolation uncertainties for SV-bas
are, of course, much smaller than those for SV-min because the
uncertainty in NMP has been fixed by choice. The strong reduction of
the uncertainties confirms once more the close relation between GR and
NMP.

The force SV-tls is fitted as SV-bas, but with the tensor spin-orbit
term switched on. There are only small changes as compared to SV-bas.
These subtle shell effects seem to have an only secondary influence on
GR.

The four lowest entries of figure \ref{fig:vary-GR} shows results from
a few conventionally used Skyrme forces. The variation of the
predictions is large confirming once more that GR are only loosely
determined by ground-state fits and that explicit adjustment is
needed for satisfying performance. \PGRcomm{Again, the GDR is
not well described by any one of the four traditional forces.
That does also hold for the force SGII which was developed originally
for GR \cite{Gia81a}, and for which we obtain the GDR peak at 12.6 MeV,
well within the results from other forces.
}

\subsubsection{GDR in the light nucleus $^{16}$O}

%
The rightmost column in  figure \ref{fig:vary-GR}
}
collects results for the peak position of the GDR in $^{16}$O.  Three
of the four NMP show strong effects ($\kappa$, $m^*/m$, and
$a_\mathrm{sym}$).  The isovector chains along $a_\mathrm{sym}$ and
$\kappa$ show the same trends which are at first glance natural in
that the $E_\mathrm{GDR}(^{16}\mathrm{O})$ peak moves up with
increasing $E_\mathrm{GDR}(^{208}\mathrm{Pb})$.
The sensitivity to variation of $m^*/m$ which is not present in
$^{208}$Pb shows that the GDR in $^{16}$O is more sensitive to shell
effects than in $^{208}$Pb. But these are all comparatively moderate
effects. The comparison with the average experimental peak reveals a
disaster.  All results stay far below the goal. We have checked all
conceivable variations within the energy-density functional
(\ref{eq:enfun}), in earlier investigations \cite{Rei99a} and in the
course of the present survey, and did not find any way to come
approximately close to the experimental
$E_\mathrm{GDR}(^{16}\mathrm{O})$ without dramatic sacrifices on the
quality of the ground-state description. 
\PGRcomm{Note that also none of the conventional Skyrme forces
is able  to reach approximately the wanted peak frequency.
}
We conclude that there is no
way to achieve a satisfying description of the GDR throughout all
nuclei with the functional (\ref{eq:enfun}). There is an urgent need
for a thorough investigation of that case.

\subsection{Fixing open options of the model}
\label{sec:options}

The SHF functional (\ref{eq:enfun}) leaves a few options open
concerning the spin-orbit model and the pairing functional. It is
worth to check whether the $\chi^2$ measure can help deciding about
preferred choices. To that end, we start from SV-bas and vary each one
of these options separately. 
\PGRcomm{
That variation proceeds similar to the
variation of NMP. We fix the selected parameter at a wanted value and
refit the other force parameters again by minimization of $\chi^2$ with
additionally constraining the NMP to the values as used in SV-bas.  We
so to say produce variants of SV-bas with one more parameter (either
$\rho_\mathrm{pair}$ or $\eta_\mathrm{tensor-ls}$ or $b'_4$) fixed.
}

\PGRcomm{
\subsubsection{Pairing model: DI versus DDDI}
}

\begin{figure}
\begin{center}
\includegraphics[width=7.8cm]{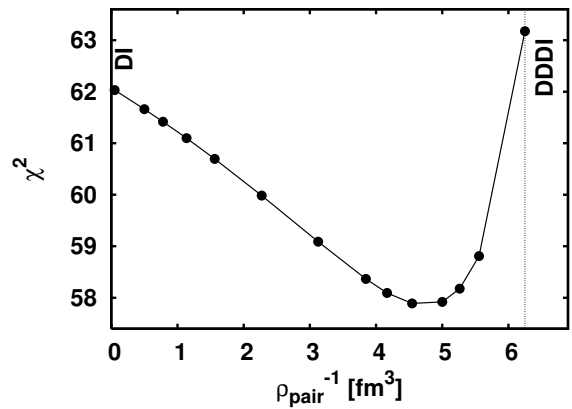}
\end{center}
\caption{\label{fig:scan-rho0pr}
Total quality measure $\chi^2$ as function of the
inverse density-switching parameter $\rho_\mathrm{pair}^{-1}$
in the variable DDDI functional.  The DI limit relates
to  $\rho_\mathrm{pair}^{-1}=0$. The standard DDDI parameter
$\rho_\mathrm{pair}=0.16\,\mathrm{fm}^{-3}$ is indicated
by a vertical dotted line.
}
\end{figure}
Figure \ref{fig:scan-rho0pr} shows the $\chi^2$ for variation of the
cutoff density $\rho_\mathrm{pair}$ in the DDDI functional
(\ref{eq:ep2}). The results are drawn versus
$\rho_\mathrm{pair}^{-1}$ because that 
{provides a better scale. The limits are:}
$\rho_\mathrm{pair}^{-1}\longrightarrow 0$ leads back to DI
while
$\rho_\mathrm{pair}^{-1}=0.16^{-1}\,\mathrm{fm}^3=6.25\,\mathrm{fm}^3$
is the typical DDDI value. The minimum $\chi^2$ obviously lies
between these two limits and the gain is considerable. We thus
decide to use $\rho_\mathrm{pair}$ as a free parameter of the pairing
model.

\PGRcomm{
\subsubsection{The tensor spin-orbit term}
}

\begin{figure}
\begin{center}
\includegraphics[width=7.8cm]{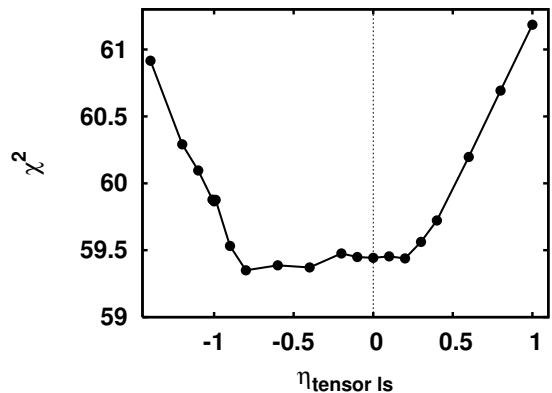}
\end{center}
\caption{\label{fig:chi-riftm}
Total quality measure $\chi^2$ as function of the
tensor-spin-orbit factor $\eta_\mathrm{tensor-ls}$.
The value  $\eta_\mathrm{tensor-ls}=0$ 
(indicated by a vertical dotted line) is associated with
ignoring tensor l*s coupling and   $\eta_\mathrm{tensor-ls}=1$
with the full term as derived from a zero-range kinetic interaction.
}
\end{figure}
The spin-orbit model leaves the option to include the tensor term
$\propto\mathbf{J}^2$ stemming from the kinetic interaction. Many
parameterizations ignore that term.  A recent compilation explored the
impact of tensor spin-orbit with flexible weight \cite{Ben07a} without
finding clear signatures {for optimum values}.  We have
introduced a continuous switch factor $\eta_\mathrm{tls}$ to allow
inclusion as tunable parameter. Figure \ref{fig:chi-riftm} shows the
result of a variation of $\eta_\mathrm{tls}$. There is a steep
increase in $\chi^2$ from $\eta_\mathrm{tls}=0$ to
$\eta_\mathrm{tls}=1$ and an almost flat landscape extending towards
moderately negative values of $\eta_\mathrm{tls}$. The extremely
shallow {$\chi^2$} landscape will make fits extremely
cumbersome {because there is no drive to a clear minimum} and
negative values {for the tensor switch factor seem} a bit
unorthodox. We thus decide to freeze the option $\eta_\mathrm{tls}=0$,
i.e. omitting tensor spin-orbit, and to consider $\eta_\mathrm{tls}=1$
(= full tensor spin-orbit) occasionally as a separate option.

\PGRcomm{
\subsubsection{Isovector spin-orbit term}
}

\begin{figure}
\begin{center}
\includegraphics[width=7.8cm]{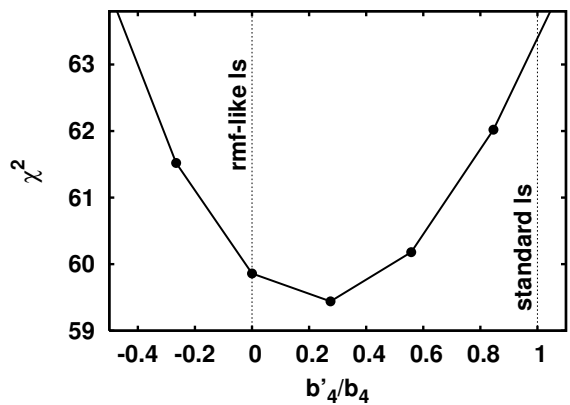}
\end{center}
\caption{\label{fig:chi-b4p}
Dependence of the
total quality measure $\chi^2$ on the 
ratio between isovector l*s coupling $b'_4$ and
isoscalar l*s coupling $b_4=t_4/2$.
%
}
\end{figure}
The other open point concerns isovector spin-orbit coupling.  That is
{described} by the parameter $b'_4$ in the Skyrme functional
(\ref{eq:enfun}). A fixed isovector fraction with $b'_4=t_4/2$ is the
standard spin-orbit model in conventional SHF functionals.  A
non-relativistic limit from the RMF suggests $b'_4\approx 0$ as the
appropriate choice \cite{Rei95a,Sul07a}. Figure \ref{fig:chi-b4p}
shows the $\chi^2$ as function of the ratio $2b'_4/t_4$. A value of
zero corresponds to the RMF preference and one to the standard
spin-orbit model of SHF. There is a considerable sensitivity. The
minimum comes close to the RMF situation for the present test
cases. However, the 
\PGRnew{optimal $b'_4/b_4$ depends somewhat on the NMP} 
constraints (e.g., effective mass, sum rule enhancement),
\PGRnew{as seen in table \ref{tab:forces}}. We thus
decide to consider $b'_4$ as a freely fitted parameter of the model.

\subsection{Performance for other observables}

In the following, we will explore the effect of variation of NMP on
various detailed observables. 
\PGRcomm{It often happens that only one NMP shows significant effects
on a given observable. In such a case,}
we will show only the one most relevant variation.

\subsubsection{Neutron skin in $^{208}$Pb}

\begin{figure}
\begin{center}
\includegraphics[width=8.4cm]{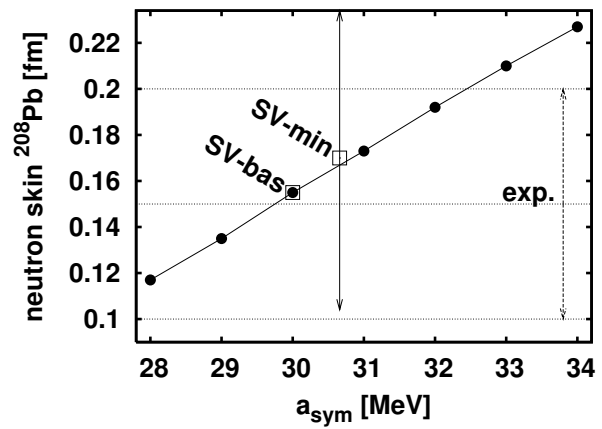}
\end{center}
\caption{\label{fig:vary-skin}
The neutron skin, $r_p-r_n$, in $^{208}$Pb 
as computed by the parameterizations with varied symmetry energy
$a_\mathrm{sym}$. The experimental value from \cite{Cla03}
is drawn as a faint horizontal line together with assumed errors of
$\pm 0.05$ fm. 
The set SV-min is shown together with the extrapolation  errors.
The extrapolation errors for SV-bas are smaller than the symbol.
}
\end{figure}
A conceptually simple observable is the neutron radius. It complements
the radius information gained from the charge form factor.
Unfortunately, its experimental determination is model dependent
because the strong interaction is involved \cite{Bat89aER}.  With that
precaution, we consider the neutron radius in $^{208}$Pb from
\cite{Cla03}. We express it as neutron skin, i.e. as difference
between neutron and proton radius, and take as experimental reference
value $r_n-r_p=0.15$ fm.
The neutron skin turns out to depend exclusively on the symmetry
energy $a_\mathrm{sym}$ while all other NMP have no effect at all. The
trend is shown in figure \ref{fig:vary-skin}. One may
even add a result from any other force into that plot and all would
line up nicely on the given slope \cite{Rei99a,Sto07aR}. The minimal
fit SV-min agrees with the data within the extrapolation error.  On
the other hand, one has to take into account that the experimental
neutron radius is extracted employing model analysis of data
\cite{Cla03}. Some uncertainty \PGRnew{is} associated with that
result. We have adopted an experimental uncertainty of 0.05 fm, which
still leaves a huge degree of freedom for $a_\mathrm{sym}$.  Only the
conventional RMF parameterizations with their rather large
$a_\mathrm{sym}>34$ MeV can safely be excluded \cite{Sto07aR}.  It is
highly desirable to have more reliable data for the neutron skin in heavy,
neutron-rich, nuclei. That would provide direct access to the symmetry
energy.

\subsubsection{The isotope shift from $^{208}$Pb to $^{214}$Pb}

\begin{figure}
\begin{center}
\includegraphics[width=8.4cm]{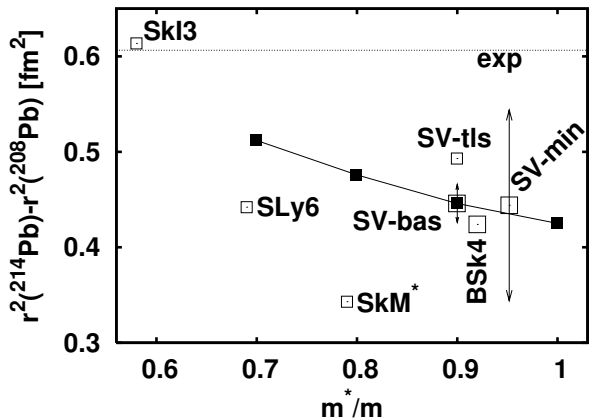}
\end{center}
\caption{\label{fig:vary-isoshift}
The isotopic shift, $r^2(^{214}\mathrm{Pb})-r^2(^{208}\mathrm{Pb})$,
of charge radii 
as computed by the parameterizations with varied effective mass
$m^*/m$ (full squares).
The experimental value is drawn as faint horizontal line \cite{Ott89aER}.
Results from some standard parameterizations are
drawn as open squares and labeled with their names in the literature.
The parameterization SV-tls is a new fit with constraints as SV-bas,
but using the full tensor spin-orbit, i.e. $\eta_\mathrm{tls}=1$,
see the SHF functional (\ref{eq:enfun}).
\PGRnew{Extrapolation errors are indicated by arrows for SV-bas and SV-min.}
}
\end{figure}

Another observable related to nuclear shape is provided by the isotope shifts of
charge radii. These are well accessible by optical methods
\cite{Ott89aER} and differences of radii put weight on aspects
complementing information from radii as such.  Particularly
interesting here is the shift
$r^2(^{214}\mathrm{Pb})-r^2(^{208}\mathrm{Pb})$ which is usually not
well described in SHF \PGRnew{with the traditional form for
the spin-orbit term} {but very well} in RMF \cite{Sha95a,Rei95a}.  We
will look at that observable with respect to the experimental result
$r^2(^{214}\mathrm{Pb})-r^2(^{208}\mathrm{Pb})=0.6085\,\mathrm{fm}^2$
\cite{Ott89aER}.
Figure \ref{fig:vary-isoshift} collects results for that isotope
shift. From the four NMP, $m^*/m$ shows the strongest effect, although
the other NMP are not totally ignorable, the forces with varying $K$
yield a variation of 0.02 fm$^2$, $a_\mathrm{sym}$ spans 0.05 fm$^2$,
and $\kappa$ 0.03 fm$^2$. All these variations stay a factor of two
below the strongest one for $m^*/m$. The non-negligible
extrapolation error of SV-bas indicates that there are also other
ingredients of the force which influence the isotopic shift. The
isovector spin-orbit force does so by construction \cite{Rei95a} and
the density-dependence of the pairing force is supposed to also have
some effect \cite{Taj93b}.  The difference between SV-tls and SV-bas
indicates the large impact of tensor spin-orbit. \PGRcomm{The
present study using systematic variation of NMP, however, reveals that} the
effective mass still has the leading influence. It was the low $m^*/m$
of the older generation of forces which \PGRcomm{enhanced the impact
of the isovector spin-orbit term with $b'_4$ such that it} lead to the
success of SkI3 and SkI4 in adjusting the isotopic shift
\cite{Rei95a}. The present \PGRnew{preference of high effective masses
leaves} little chance to come so close to the data as can be seen from
the error bars on SV-min.
\PGRnew{This calls for new investigations on that subject.}

\subsubsection{Neutron level sequence in $^{132}$Sn}

\begin{figure}
\begin{center}
\includegraphics[width=8.4cm]{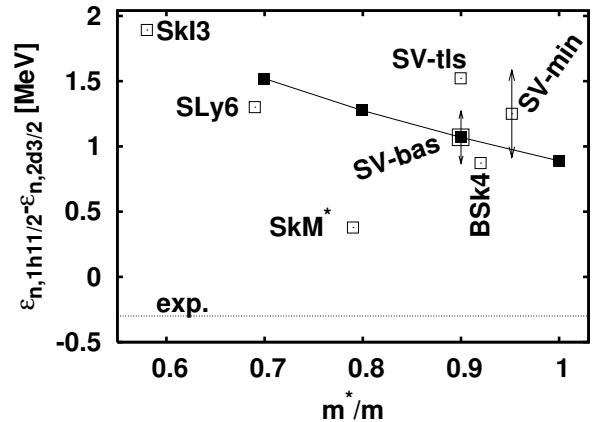}
\end{center}
\caption{\label{fig:vary-levSn}
The energy difference between two occupied neutron levels near the
Fermi surface in $^{132}$Sn,
$\varepsilon_\mathrm{n,1h11/2}-\varepsilon_\mathrm{n,2d3/2}$,
for the chain of varying 
effective mass and some other Skyrme parameterizations as indicated.
The experimental value is indicated by a faint horizontal line
\cite{Kin97aER}.
\PGRnew{Extrapolation errors are indicated by arrows for SV-bas and SV-min.}
}
\end{figure}
The single-particle energies are a subtle observable in connection
with mean-field models. Density functional theory does not give a
guarantee that they are correctly described \cite{Dre90aB} and indeed
they are hampered by the self-interaction error. That error, however,
leaves the energy differences untouched \cite{Leg02} and it is anyway
less dramatic in nuclei. The robustness of single-particle energy
differences has also been confirmed in the nuclear context
\cite{Rut98a,Ber80a}. Moreover, the detailed level structure plays a
crucial role for the properties of SHE \cite{Ben01a}. It is thus
important to investigate the performance of mean-field models in that
respect. Simple test cases can only be doubly magic nuclei to avoid
perturbations from pairing and polarization effects.  The levels are
usually described fairly well in $^{208}$Pb to the extent that at
least the level ordering is usually correctly reproduced and often
also the detailed energy differences \cite{Ben03aR}. However, a
particularly obnoxious case is the sequence of occupied neutron levels
in $^{132}$Sn. The experimental ground state of $^{131}$Sn has a spin
$3/2^+$ while all mean field models predict $11/2^-$ spin. We take the
energy difference between these two states, 
$\varepsilon_\mathrm{n,1h11/2}-\varepsilon_\mathrm{n,2d3/2}$,
as one number characterizing the sequence. 
Figure \ref{fig:vary-levSn} shows the results for the chain with
varied $m^*/m$. It is not surprising that this variation has most
effect because $m^*/m$ is closely related to shell structure.  The
next sensitive NMP is $\kappa$ (not shown here) which is related to
the isovector effective mass. The other parameterizations in the
figure \PGRcomm{gather between 1--2} MeV and all theoretical results
(including those from the RMF \cite{Rut98a,Ben03aR}) are far off the
experimental value.

The origin of this discrepancy is not clear. It may be related to a
peculiarity of the high angular momentum state involved in that
difference, either that its spin-orbit splitting is underestimated or
that the mean position of high spatial angular momenta is too high. It
would require extensions of the SHF functional to cure one of these
two features. But it could also be a problem with the interpretation
of the excitation spectrum in $^{131}$Sn as neutron levels in
$^{132}$Sn. What has not yet been checked so far is the influence of
particle-core coupling in $^{132}$Sn. The problem of the
level sequence calls for further thorough investigations.

A similar discrepancy is seen for the energy difference between the
proton $1h11/2$ and $2g7/2$ levels in Sn isotopes \cite{Sch04a} where
all mean field models yet fail to reproduce the isotopic trend of the
splitting. Two states with high angular momentum are involved. This
suggests that an insufficient description of spin-orbit splitting is
the more likely source of trouble. But again, this has yet to be
investigated in more detail.

\subsubsection{Extrapolations to neutron-rich Sn Isotopes}

\begin{figure}
\begin{center}
\includegraphics[width=7.4cm]{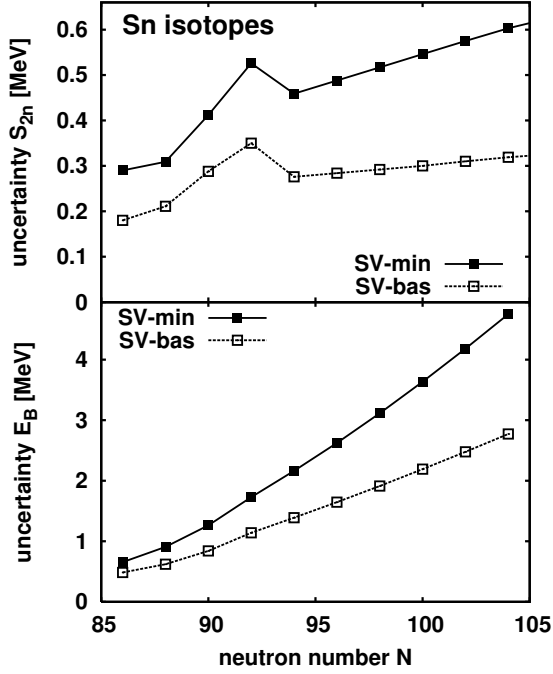}
\end{center}
\caption{\label{fig:extrap-chain}
Performance for extrapolations to extremely neutron-rich
Sn isotopes for the two parameterizations SV-min and SV-bas.
Lower: Predicted extrapolation uncertainties for binding
energies. 
Upper: Predicted extrapolation uncertainties for two-neutron
separation energies $S_{2n}$.
}
\end{figure}
Astrophysical applications for $r$-process nuclei involve
extrapolations deep into the regime of neutron-rich isotopes. Figure
\ref{fig:extrap-chain} shows the extrapolation errors as estimated
from the least-squares techniques for binding energies and two-neutron
separation energies along the chain of exotic Sn isotopes.
The lower panel for binding energies shows a systematic growth of the
uncertainty when moving away from the valley of stability. Note that
the freezing of crucial NMP by \PGRnew{tuning} of giant resonances in SV-bas
reduces the uncertainty by a factor of two.
The upper panel for two-neutron separation energies basically behaves
similarly, but the errors are much smaller than for the energies as
such.
Differences of energies probe the response properties and these
are obviously a bit more robust. Note also the peak at neutron
number $N=92$. There is a small sub-shell closure which is particularly
sensitive to shell structure. The increase of uncertainty here
indicates that 
\PGRnew{an aspect of shell structure comes into play which is not
so well determined as the general trends.}


\PGRcomm{

\subsubsection{Fission barriers in $^{236}$U}

\begin{figure}
\begin{center}
\includegraphics[height=7.4cm,angle=-90]{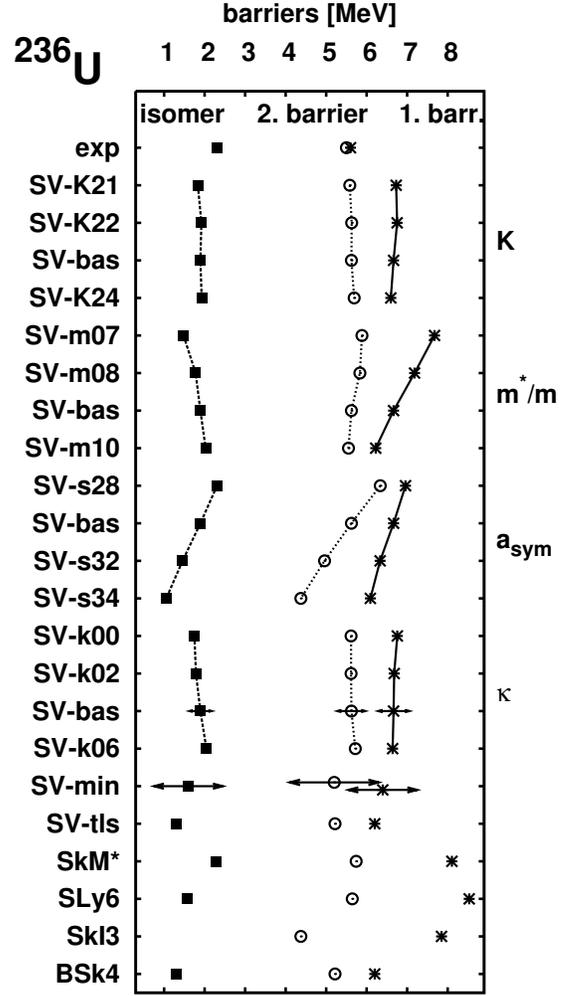}
\end{center}
\caption{\label{fig:vary-barrier-U-3}
Axially symmetric fission barriers (star = first barrier, open circle
= second barrier) and position of the isomeric minimum (filled square)
in $^{236}$U for the variety of parameterizations as indicated.  The
first barrier Extrapolation errors are shown for SV-bas and SV-min.
The errors for SV-bas are representative for all forces with
constrained NMP. The experimental values are taken from
\cite{Bjo80aER,Mam98a}.
}
\end{figure}
Fission of actinides is a crucial nuclear \PGRnew{process}.  It is usually
characterized in terms of fission barriers.  We deduce the barriers
from quadrupole-constrained SHF calculations considering axially
symmetric shapes and allowing for reflection asymmetry. 
\PGRnew{A series of constrained calculations for a broad range
of deformations}
spans the fission path and the associated
deformation energy surface.  All states along the path are well
deformed. We thus \PGRnew{compute} the energy with the 
correction from angular momentum projection in GOA, as explained in section
\ref{sec:framework}. That energy curve
for actinides  displays the typical double-humped barrier \cite{Bjo80aER} with
one isomeric minimum. From that we read off two barrier heights and the
isomeric energy. It is to be noted that the first barrier tends to go
through triaxial shapes, not accounted for here. A further lowering by
0.5--2 MeV can be expected from triaxiality
\cite{Cwi96a,Ben98a}.

Figure \ref{fig:vary-barrier-U-3} shows results for fission isomer and
fission barrier for $^{236}$U as a typical example for an
actinide. The values fit generally well to the experimental data,
particularly when considering some triaxial lowering for the first
barrier. The strongest influence from NMP on the first barrier comes
from the symmetry energy $a_\mathrm{sym}$, some effect is added from
$m^*/m$, and $K$ as well as $\kappa$ remain basically inert. The order
is reversed for the second barrier where $m^*/m$ has largest impact
and $a_\mathrm{sym}$ is secondary.
\PGRnew{These two NMP are the crucial handles on fission properties
including the isomer.}
This reflects that fission emerges from a subtle interplay of bulk
properties (here $a_\mathrm{sym}$) and shell effects (here $m^*/m$).
The impact of shell effects through the spin-orbit force is seen also
in the step from SV-bas to SV-tls which yields a small, but
non-negligible, lowering of the barriers.

It is interesting to  note that the variation of results for the
conventional forces
is larger  than the variation  within all  SV forces. That indicates 
a large sensitivity of fission properties to fitting strategies.
}
%



\PGRcomm{
\subsubsection{A known SHE:  $^{264}$Hs}
\label{sec:SHE}
\begin{figure}
\begin{center}
\includegraphics[height=7.4cm,angle=-90]{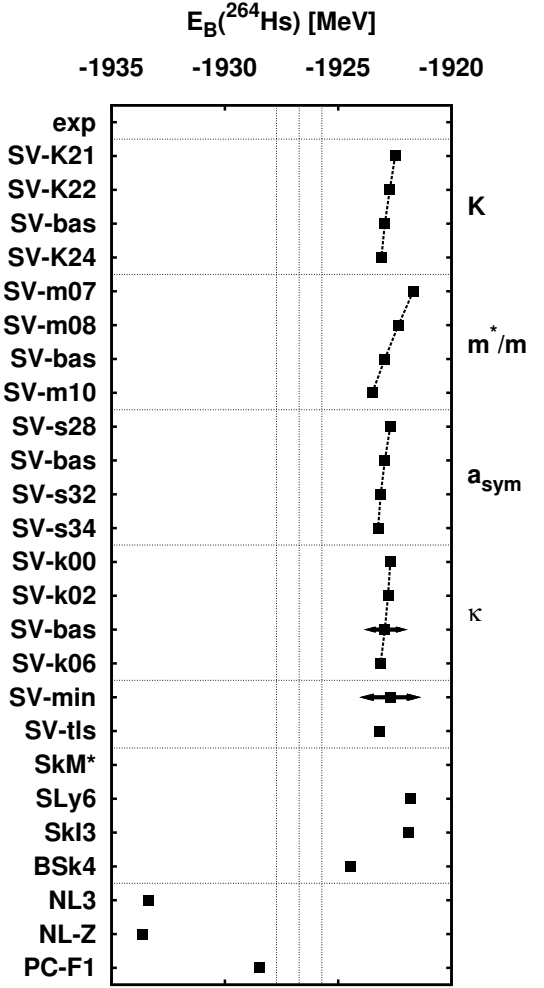}
\end{center}
\caption{\label{fig:vary-EHs}
The binding energy  of $^{264}$Hs for the variety
of parameterizations as indicated. The lowest three entries
show results from relativistic mean-field models
\cite{Bue98a,Bue02a}. 
Extrapolation errors are shown for SV-bas and SV-min.
The errors for SV-bas are representative for all forces
with constrained NMP.
}
\end{figure}
Some SHE have already been produced such that data are available for
probing the predictive power directly. We consider here $^{264}$Hs as
sample.  The nucleus has a well deformed ground state. In fact, it
belongs to an island of deformed shell closures \cite{Sob89a,Cwi96a}.
We discussed it, amongst others, already in connection with figures
\ref{fig:SV-min-energies} and \ref{fig:SV-def-energies-2}.  Here, we
continue with more detailed variations.  Figure \ref{fig:vary-EHs}
shows the binding energy of $^{264}$Hs for all forces in our sample.
The test case embraces extrapolation to SHE together with deformation
effects. It is thus very sensitive and we find that all NMP
have some influence, the strongest coming from  $m^*/m$.
But none of the allowed variations of NMP brings the result in any way
close to the experimental value. The same holds true for the
conventional Skyrme forces. That feature was already observed
in \cite{Bue98a}
and it was also  found that relativistic mean field models 
behave quite differently. Thus we  have added a few results from
relativistic models, NL-Z \cite{Ruf88a},  NL3 \cite{Lal97a}, and
PC-F1 \cite{Bue02a}. These tend to overbinding where SHF generally
yields underbinding.  The result indicates a deep rooted structural
difference between these two classes of models, and possibly missing
terms in both.
}

\begin{figure}
\begin{center}
\includegraphics[height=7.4cm,angle=-90]{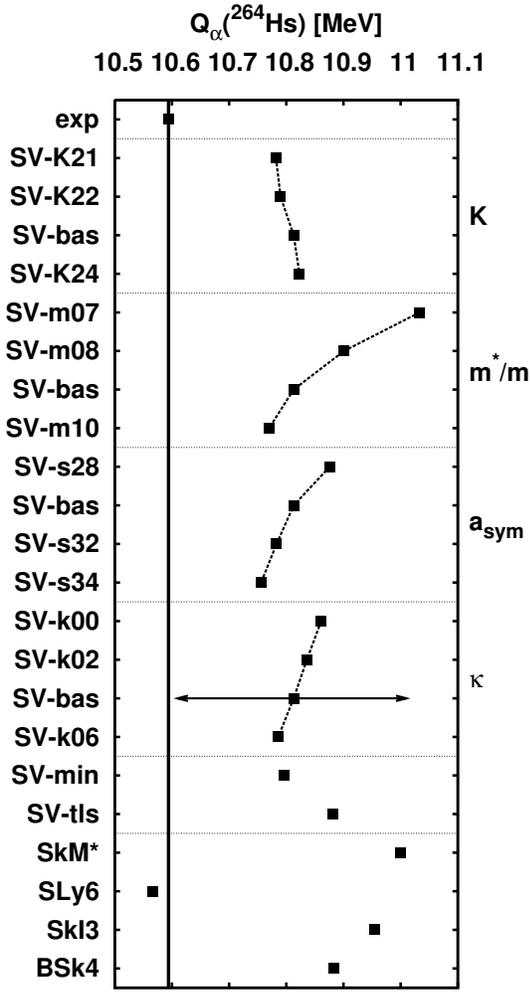}
\end{center}
\caption{\label{fig:vary-Qalpha-Hs}
The $Q_\alpha$ value of $^{264}$Hs for the variety
of parameterizations as indicated.
Extrapolation errors are shown for SV-bas.
Those for SV-min exceed the bounds of the plot.
}
\end{figure}
Reactions in the landscape of SHE are not so much determined by
binding energies as such. Differences of binding energies are more
crucial, \PGRnew{particularly the $Q_\alpha$ value which characterizes
$\alpha$-decay}. Figure \ref{fig:vary-Qalpha-Hs} shows $Q_\alpha$ for
$^{264}$Hs in comparison to the experimental value.  All forces
provide rather nice agreement with the experimental value. There is
general trend to about 0.2 MeV underestimation.  This may be due to
the fact that the daughter nucleus $Z=106$ is softer than the parent
$Z=108$ such that some additional correlation effect may come into
play. Nevertheless, the agreement is much better than what one could
have expected from the variations in the binding energies (see figure
\ref{fig:vary-EHs}). That indicates that energy differences can be
predicted more safely.

\PGRcomm{
\subsubsection{At the upper end a spherical SHE:  {$Z,N=(120,182)$}}
\label{sec:SHE2}

Further up, there are very interesting, yet unmeasured, SHE. 
In that regime, we
consider as test
case element {$Z,N=(120,182)$}, at the upper edge of experimental
feasibility in the regime of the spherical valley of stability. It
resides on an upward extension of an $\alpha$-decay chain recently
detected \cite{Oga99cE,Oga00aE,Oga01aE}.
 This is a
spherical nucleus due to the proton shell closure at $Z=120$ and a
\PGRnew{a low level density for the neutrons at their Fermi energy}
\cite{Ben99a,Kru00a,Ber01a}.
}
We consider for that element the binding energy
$E_B$ as such, the $Q_\alpha$ value 
\begin{equation}
  Q_\alpha(Z,N)
  =
  E_B(Z\!-\!2,N\!-\!2)
  +
  E_B(2,2)
  -
  E_B(Z,N)
\label{eq:qalpha}
\end{equation}
which characterizes $\alpha$-decay, and the fission barrier
$B_\mathrm{fis}$. The latter quantity is deduced from computing the
potential-energy surface along the axially symmetric quadrupole
deformation path, for details see \cite{Bue04a}, including again the
correction from angular momentum projection. Most SHE have only a single
fission barrier.

\begin{figure*}
\begin{center}
\includegraphics[height=17cm,angle=-90]{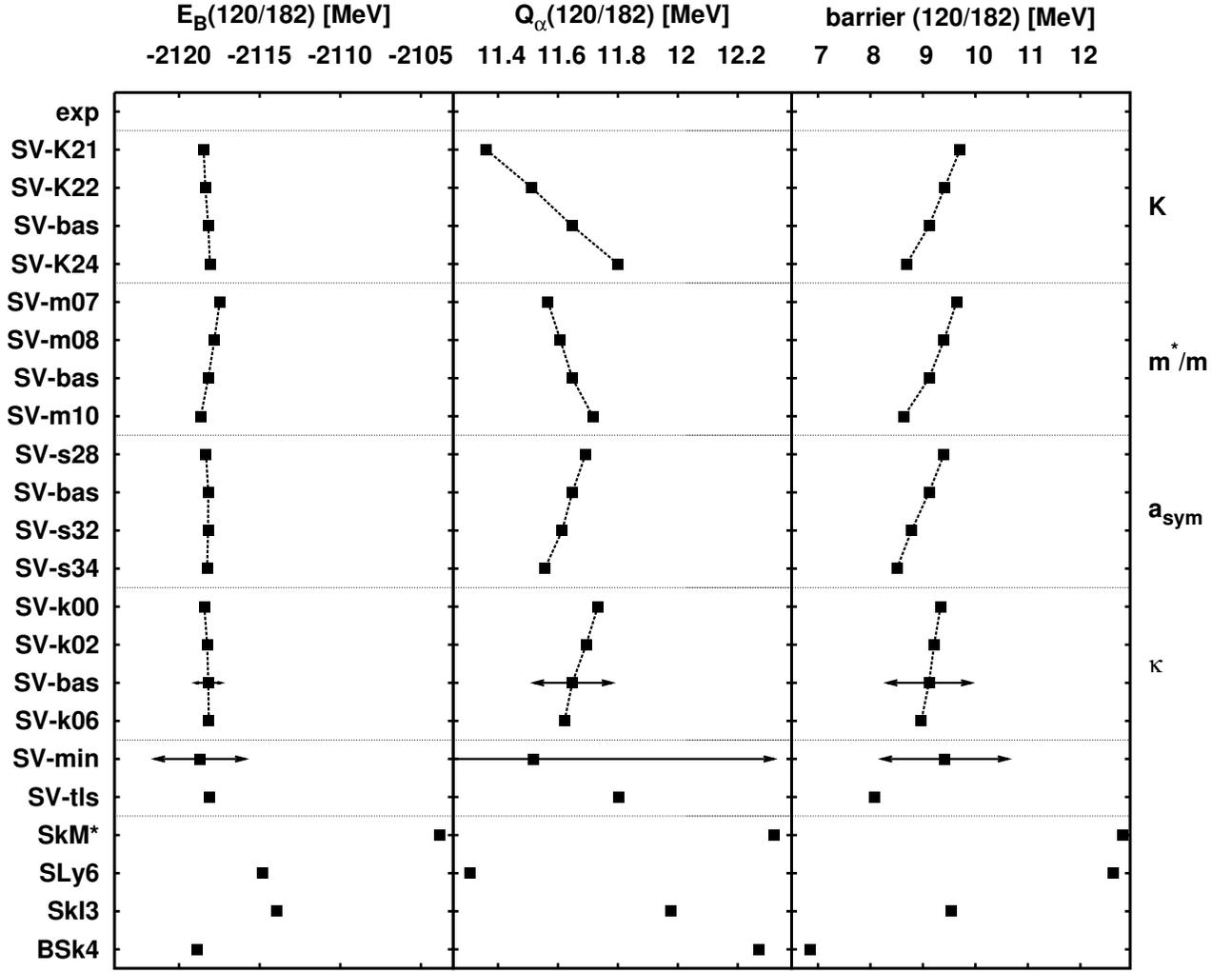}
\end{center}
\caption{\label{fig:vary-she-3}
Results of the various forces for the super-heavy element
$(Z/N)=(120/182)$. Left column: total binding energy.
Middle column: $Q_\alpha$ value. Right column: fission barrier.
Extrapolation errors are shown for SV-bas and SV-min.
The errors for SV-bas are representative for all forces
with constrained NMP.
}
\end{figure*}
\PGRcomm{
Figure \ref{fig:vary-she-3} shows the results 
for the hypothetical nucleus (Z/N)=(120/182). We
concentrate first on the binding energy (left column).
}
The span of predictions from
conventional parameterizations is huge. The force SkM$^*$ is far off
all other results which could be explained by two reasons: First, it
is the oldest force in the sample and it was adjusted on a smaller
data base available at that time \cite{Bar82a}.  Second, and
probably more important, SkM$^*$ is the only force in the sample which
uses a different recipe for the c.m. correction, namely to take only
the diagonal elements of the $\hat{P}_\mathrm{cm}$ in eq.
(\ref{eq:cmfull}) which has dramatic consequences for the
extrapolation to SHE \cite{Ben00b}. The more recent forces are grouped
\PGRcomm{somewhat
better together, and the whole set of SV forces shows comparatively
little variation.
The extrapolation error of SV-min 
is still smaller than the difference to conventional forces
and}
SV-bas comes down to an uncertainty of $\pm 0.8$ MeV. The difference
in the extrapolation errors between SV-min and SV-bas shows the
influence of the loosely fixed NMP. A better determination of NMP
through GR does also improve the predictive value in the regime
of exotic nuclei.
%

%

%
\PGRcomm{ The right column of figure \ref{fig:vary-she-3}
shows the fission barriers for the SHE 120/182. 
These}
are sensitive to all details of a
parameterization {due to the subtle interplay of bulk properties
(surface tension, Coulomb pressure) and shell effects}. 
All NMP yield 1--2 MeV variation of the result. The sizeable
extrapolation errors of about 0.8 MeV for SV-bas show that there are
other effects also at work. The fission barrier is determined by bulk
properties (i.e. NMP) as well as shell effects \cite{Bue04a}. The
latter can be seen, e.g., by the large effect of the tensor spin-orbit
force, see the difference between SV-bas and SV-tls. In spite of all
these sensitivities, the more conservative extrapolation error of
SV-min indicates that the predictions can be taken within about 1.2
MeV reliability. It is comforting to see that most of the conventional
parameterizations stay also within these bounds. Thus we see that SHF
notoriously predicts the fission barrier for 120/182 around 6
MeV 
\PGRcomm{with angular momentum correction and 7-8  MeV without.}
That remains in great contrast to the RMF where all predictions 
come out much lower \cite{Bue04a}.

%
There is a competition between $\alpha$-decay and fission in the decay
channels of SHE. 
\PGRcomm{ The middle column of figure \ref{fig:vary-she-3}}
shows the $Q_\alpha$
value for the SHE 120/182. This quantity is directly deduced from a
difference of binding energies. One would expect similar trends as for
the binding energies (left column), but differences
can suppress one trend and amplify another. That is what happens for
the  $Q_\alpha$. The trend with $m^*/m$ looses importance and the
incompressibility $K$ acquires more weight, but the variations are
generally  much smaller than for the fission barriers. Except for
SkM$^*$, all forces agree within $\pm 0.4$ MeV. The fact that the
extrapolation error for SV-bas is much smaller than for SV-min
indicates that NMP are the leading determinators, in contrast to
fission barriers where shell effects have much larger influence.

\section{Conclusion}

In this paper, we have performed a survey of the phenomenological
adjustment of the free parameters of the Skyrme-Hartree-Fock (SHF)
\PGRcomm{energy functional}. The input data for the fits are taken
from basic nuclear ground-state properties: energies, charge radii,
surface thickness, selected odd-even staggering of energies, and some
spin-orbit splittings.  These data and appropriate nuclei are selected
carefully such that they have small correlation effects. Thereby we
consider only the strongly fluctuating correlations from low-lying
collective quadrupole states. The smoothly varying correlations from
higher excitations (two-body collisions and resonances) are supposed
to be incorporated effectively in the SHF functional. 
\PGRnew{The investigation of correlation effects led to a selection of
semi-magic fit nuclei with large extension along isotonic chains but
surprisingly short isotopic chains. All selected nuclei are spherical
to avoid ambiguities from the handling of deformed minima
(angular momentum projection).}
A quality
measure $\chi^2$ is built from summing the squared deviations from the
data with appropriate adopted error weights. The parameters of the SHF
functional are optimized by least-squares fits with respect to the
selected sample of data.
\PGRnew{
The emerging r.m.s. errors in the fit observables stay well (by a factor
two) below the adopted input errors from correlation effects which
shows that the SHF model has some flexibility to incorporate part of
correlations from the low-lying excitations.
}

\PGRcomm{It is found that \PGRnew{the $\chi^2$} minimization leaves
some  freedom in nuclear bulk properties. In order to explore
the space of well-fitted forces thus becoming available, we have performed
a systematic  variation of bulk  properties, characterized in terms
of nuclear matter properties (NMP). This is achieved by 
adding to the selected ground-state data
constraints on four NMP:}
incompressibility $K$, effective mass $m^*/m$, symmetry energy
$a_\mathrm{sym}$, and sum rule enhancement factor $\kappa$. These NMP
are varied systematically to produce a set of forces with different
properties, all having about the same high quality concerning the
nuclear ground-state properties in the fit data. The set of
parameterizations thus obtained was used for a thorough investigation
of the predictive power of the SHF model by looking at the results for
several detailed observables in stable and super-heavy nuclei.


We have used the quality measure $\chi^2$ to check some open options
of the present ansatz for the SHF functional. There used to be the
decision between volume pairing which corresponds to a simple
zero-range pairing force and surface pairing which augments that with
a density dependence such that pairing is basically switched off for
densities near bulk equilibrium. We have allowed a flexible
switching
\PGRcomm{with a density parameter between bulk equilibrium (surface
pairing) and
infinity (volume pairing).}
It turns out that the optimum choice is just between these
extremes.
The SHF functional allows a free choice for the isovector spin-orbit
force. A free variation is found to be advantageous as compared to
the standard choice which \PGRnew{is} linked to the isoscalar value or to the
choice of zero isovector term deduced from the RMF. 
There is, furthermore, the choice to include the tensor spin-orbit
term which is related to the kinetic zero-range two-body interaction.
The quality measure prefers a model without tensor spin-orbit.
All three decisions, however, are related to changes in the total
$\chi^2$ by about 5\%.


We have looked at the separate contributions of an observable
(e.g., energy) to the total $\chi^2$. The trends of these contributions
with systematically varied NMP pull often in different
directions. This means that the choice of relative weights of one
observable with respect to another decides on the final result.  The
trends are weak and thus small changes in weight can cause large
drifts in final NMP of a least-squares minimum.  This is a quite
undesirable element of arbitrariness in the adjustment procedure. On the
other hand, the weakness of the trends means that there are several
features not so sharply determined from ground-state fits. We exploit
the freedom to additionally adjust the peak positions of three
decisive giant resonances in $^{208}$Pb: the isoscalar giant monopole
resonance (GMR), the isoscalar giant quadrupole resonance (GQR), and
the isovector giant dipole resonance (GDR). The GMR fixes the
incompressibility and the GQR the effective mass.
\PGRnew{Both resonances can
be well described by the model}. The GDR is less conclusive in two
respects. First, it is sensitive to two NMP, the symmetry energy and
the sum rule enhancement factor. And second, it does not yet allow
fully satisfying tuning due to the strong fragmentation pattern.
Moreover, we find that the GDR in $^{16}$O is far off the experimental
peak position for all reasonable SHF parameterizations. The present
forms of the SHF functional do not yet allow a proper description of
the GDR throughout all nuclei. This case of the GDR still requires
further investigation.

In summary, we have built one force SV--min with minimal $\chi^2$ from
an unrestricted fit, another force SV-bas from NMP constrained fits
with GR fine-tuning, a series of forces with systematically varied
NMP, and finally a variant of SV-bas with tensor spin-orbit term.
Using this set of trial forces, we  have investigated 
\PGRcomm{the performance for other nuclei and}
the predictions for a variety of more detailed observables. 
\PGRcomm{
The interpolation to nuclei within the fitted mass range, i.e.
$A<220$, perform fairly well when including angular momentum
projection. The remaining underbinding of 1--2 MeV for soft
nuclei is resolved by vibrational correlations.
The extrapolation to heavier, deformed nuclei, however,
develops large underbinding up  to 4 MeV. The attempt to cure that by
including super-heavy elements in the fit did  not work out well.
It sacrifices too much of the quality reached for stable  nuclei.
We encounter here most probably  a  problem with an insufficient
form of the present energy functional.
}
The binding energy of spherical super-heavy nuclei is a rather robust
quantity depending mainly on the effective mass and other features
influencing shell structure. Deformed  super-heavy nuclei are more sensitive
because of the deformation energy. 
The fission barriers in super-heavy elements is more sensitive
depending on almost any feature of the force, all NMP and more detailed 
entries as, e.g.,  the tensor spin-orbit term. Nonetheless, all
predictions lie within a band of $\pm 1.2$ MeV out of about 10 MeV
which is a very comforting result for such a subtle observable.
Even more robust are the $Q_\alpha$ energies determining the rate of
$\alpha$ decay, which are most strongly influenced by the
incompressibility. This shows that differences of energies can behave
differently from the energies as such (where the effective mass was most
influential).
The neutron skin in $^{208}$Pb depends exclusively on the symmetry
energy. It would be the ideal means to determine this crucial
isovector property, but the measurement needs to be reliable and
precise. An uncertainty of 0.02 fm in the neutron radius translates to
an uncertainty of 1 MeV in the symmetry energy.  Practically all
existing SHF parameterizations comply with the presently available data
within their large uncertainties.
The isotope shift of charge radii between $^{214}$Pb and $^{208}$Pb
depends sensitively on the effective mass and other features
influencing the shell structure. The recent fits all tend to rather
large effective mass which, in turn, {raises problems with
accommodating that isotopic shift}, even when allowing full freedom in
the isovector spin-orbit term. The case, which seemed to be solved for
previous parameterizations with low effective mass, is open again and
requires further investigations.

Altogether, we find that the SHF model provides an excellent
description of nuclear bulk properties. Ground-state properties alone,
however, leave several features weakly determined. Additional
information from excitation or more detailed ground-state observables
is required to fix all aspects of the SHF model. 
At the same time, the added observables
reveal some insufficiencies of the model which call for further
investigations. Most urgent seems to be an extension of the spin-orbit
model and a better description of dipole giant resonances.

\section*{Acknowledgement}

This work was supported by BMBF under contracts no. 06 FY 159D and 06 ER
142D. We gratefully acknowledge support by the
Regional Computing Center Erlangen.

\appendix

\section{Details on data and parameters}
\label{app:details}

\begin{table*}
\begin{tabular}{|rr|rr|rr|rr|rr|rr|rr|rr|rr|}
\hline
 A &    Z &
  \multicolumn{2}{|c|}{$E_B$} &
  \multicolumn{2}{|c|}{$R_\mathrm{diffr}$} &
  \multicolumn{2}{|c|}{$\sigma$} &
  \multicolumn{2}{|c|}{$r_\mathrm{rms}$} &
  \multicolumn{2}{|c|}{$\varepsilon_{ls,p}$} &
  \multicolumn{2}{|c|}{$\varepsilon_{ls,n}$} &
  \multicolumn{2}{|c|}{$\Delta_p$} &
  \multicolumn{2}{|c|}{$\Delta_n$} \\
\hline
    &     &  
  \multicolumn{2}{|c|}{$\pm 1$ MeV} &
  \multicolumn{2}{|c|}{$\pm 0.04$ fm} &
  \multicolumn{2}{|c|}{$\pm 0.04$ fm} &
  \multicolumn{2}{|c|}{$\pm 0.02$ fm} &
  \multicolumn{2}{|c|}{$\pm 20\%$ } &
  \multicolumn{2}{|c|}{$\pm 20\%$ } &
  \multicolumn{2}{|c|}{$\pm 0.12$ MeV} &
  \multicolumn{2}{|c|}{$\pm 0.12$ MeV} \\
\hline
 16 &   8 &  -127.620 & 4 & 2.777 & 2 & 0.839 & 2 & 2.701 & 2 &  6.30 & 3 &  6.10 & 3 &      &   &      &  \\
\hline
 36 &  20 &  -281.360 & 2 &       &   &       &   &       &   &       &   &       &   &      &   &      &  \\
 38 &  20 &  -313.122 & 2 &       &   &       &   &       &   &       &   &       &   &      &   &      &  \\
 40 &  20 &  -342.051 & 3 & 3.845 & 1 & 0.978 & 1 & 3.478 & 1 &       &   &       &   &      &   &      &  \\
 42 &  20 &  -361.895 & 2 & 3.876 & 1 & 0.999 & 1 & 3.513 & 2 &       &   &       &   &      &   & 1.68 & 4\\
 44 &  20 &  -380.960 & 2 & 3.912 & 1 & 0.975 & 1 & 3.523 & 2 &       &   &       &   &      &   & 1.70 & 2\\
 46 &  20 &  -398.769 & 2 &       &   &       &   & 3.502 & 1 &       &   &       &   &      &   & 1.49 & 4\\
 48 &  20 &  -415.990 & 1 & 3.964 & 1 & 0.881 & 1 & 3.479 & 2 &       &   &       &   &      &   &      &  \\
 50 &  20 &  -427.491 & 1 &       &   &       &   & 3.523 & 9 &       &   &       &   &      &   &      &  \\
 52 &  20 &  -436.571 & 1 &       &   &       &   &       &   &       &   &       &   &      &   &      &  \\
\hline
 56 &  28 &  -483.990 & 5 &       &   &       &   & 3.750 & 9 &       &   &       &   &      &   &      &  \\
 58 &  28 &  -506.500 & 5 & 4.364 & 1 &       &   & 3.776 & 5 &       &   &       &   &      &   &      &  \\
 60 &  28 &  -526.842 & 5 & 4.396 & 1 & 0.926 & 5 & 3.818 & 5 &       &   &       &   &      &   &      &  \\
 62 &  28 &  -545.258 & 5 & 4.438 & 1 & 0.937 & 5 & 3.848 & 5 &       &   &       &   &      &   &      &  \\
 64 &  28 &  -561.755 & 5 & 4.486 & 1 & 0.916 & 2 & 3.868 & 5 &       &   &       &   &      &   &      &  \\
 68 &  28 &  -590.430 & 1 &       &   &       &   &       &   &       &   &       &   &      &   &      &  \\
\hline
100 &  50 &  -825.800 & 2 &       &   &       &   &       &   &       &   &       &   &      &   &      &  \\
108 &  50 &           &   &       &   &       &   & 4.563 & 2 &       &   &       &   &      &   &      &  \\
112 &  50 &           &   & 5.477 & 3 & 0.963 & 9 & 4.596 & 9 &       &   &       &   &      &   & 1.41 & 9\\
114 &  50 &           &   & 5.509 & 3 & 0.948 & 9 & 4.610 & 9 &       &   &       &   &      &   & 1.26 & 9\\
116 &  50 &           &   & 5.541 & 3 & 0.945 & 9 & 4.626 & 9 &       &   &       &   &      &   & 1.21 & 9\\
118 &  50 &           &   & 5.571 & 2 & 0.931 & 2 & 4.640 & 1 &       &   &       &   &      &   & 1.34 & 9\\
120 &  50 &           &   & 5.591 & 1 &       & 1 & 4.652 & 1 &       &   &       &   &      &   & 1.39 & 9\\
122 &  50 & -1035.530 & 3 & 5.628 & 1 & 0.895 & 1 & 4.663 & 1 &       &   &       &   &      &   & 1.37 & 3\\
124 &  50 & -1050.000 & 3 & 5.640 & 1 & 0.908 & 1 & 4.674 & 1 &       &   &       &   &      &   & 1.31 & 3\\
126 &  50 & -1063.890 & 2 &       &   &       &   &       &   &       &   &       &   &      &   & 1.26 & 2\\
128 &  50 & -1077.350 & 2 &       &   &       &   &       &   &       &   &       &   &      &   & 1.22 & 2\\
130 &  50 & -1090.400 & 1 &       &   &       &   &       &   &       &   &       &   &      &   & 1.17 & 3\\
132 &  50 & -1102.900 & 1 &       &   &       &   &       &   &  1.35 & 1 &  1.65 & 1 &      &   &      &  \\
134 &  50 & -1109.080 & 1 &       &   &       &   &       &   &       &   &       &   &      &   &      &  \\
\hline
198 &  82 & -1560.020 & 9 &       &   &       &   & 5.450 & 2 &       &   &       &   &      &   &      &  \\
200 &  82 & -1576.370 & 9 &       &   &       &   & 5.459 & 1 &       &   &       &   &      &   &      &  \\
202 &  82 & -1592.203 & 9 &       &   &       &   & 5.474 & 1 &       &   &       &   &      &   &      &  \\
204 &  82 & -1607.521 & 2 & 6.749 & 1 & 0.918 & 1 & 5.483 & 1 &       &   &       &   &      &   & 0.77 & 2\\
206 &  82 & -1622.340 & 1 & 6.766 & 1 & 0.921 & 1 & 5.494 & 1 &       &   &       &   &      &   & 0.59 & 3\\
208 &  82 & -1636.446 & 1 & 6.776 & 1 & 0.913 & 1 & 5.504 & 1 &  1.42 & 1 &  0.90 & 1 &      &   &      &  \\
    &     &           &   &       &   &       &   &       &   &       &   &  1.77 & 2 &      &   &      &  \\
210 &  82 & -1645.567 & 1 &       &   &       &   & 5.523 & 1 &       &   &       &   &      &   & 0.66 & 3\\
212 &  82 & -1654.525 & 1 &       &   &       &   & 5.542 & 1 &       &   &       &   &      &   &      &  \\
214 &  82 & -1663.299 & 1 &       &   &       &   & 5.559 & 1 &       &   &       &   &      &   &      &  \\
\hline
\end{tabular}
\caption{\label{tab:fitdata1}
Experimental data for the fits, part I: along isotopic chains.
Each column stands for an observable as indicated.
The second line shows the globally adopted error for each observable.
{That error is multiplied  for each observable by a further
integer weight factor which is given in the column next to the 
data value.}
}
\end{table*}
\begin{table*}
\begin{tabular}{|rr|rr|rr|rr|rr|rr|rr|rr|rr|}
\hline
 A &    Z &
  \multicolumn{2}{|c|}{$E_B$} &
  \multicolumn{2}{|c|}{$R_\mathrm{diffr}$} &
  \multicolumn{2}{|c|}{$\sigma$} &
  \multicolumn{2}{|c|}{$r_\mathrm{rms}$} &
  \multicolumn{2}{|c|}{$\varepsilon_{ls,p}$} &
  \multicolumn{2}{|c|}{$\varepsilon_{ls,n}$} &
  \multicolumn{2}{|c|}{$\Delta_p$} &
  \multicolumn{2}{|c|}{$\Delta_n$} \\
\hline
    &     &  
  \multicolumn{2}{|c|}{$\pm 1$ MeV} &
  \multicolumn{2}{|c|}{$\pm 0.04$ fm} &
  \multicolumn{2}{|c|}{$\pm 0.04$ fm} &
  \multicolumn{2}{|c|}{$\pm 0.02$ fm} &
  \multicolumn{2}{|c|}{$\pm 20\%$ } &
  \multicolumn{2}{|c|}{$\pm 20\%$ } &
  \multicolumn{2}{|c|}{$\pm 0.12$ MeV} &
  \multicolumn{2}{|c|}{$\pm 0.12$ MeV} \\
\hline
 34 &  14 &  -283.429 & 2 &       &   &       &   &       &   &       &   &       &   & 2.22 & 8 &      &  \\
 36 &  16 &  -308.714 & 2 & 3.577 & 4 & 0.994 & 4 & 3.299 & 1 &       &   &       &   & 1.52 & 8 &      &  \\
 38 &  18 &  -327.343 & 2 &       &   &       &   & 3.404 & 1 &       &   &       &   & 1.44 & 8 &      &  \\
 42 &  22 &  -346.904 & * &       &   &       &   &       &   &       &   &       &   &      &   &      &  \\
\hline
 50 &  22 &  -437.780 & 2 & 4.051 & 1 & 0.947 & 2 & 3.570 & 1 &       &   &       &   &      &   &      &  \\
 52 &  24 &  -456.345 & * & 4.173 & 1 & 0.924 & 4 & 3.642 & 2 &       &   &       &   &      &   &      &  \\
 54 &  26 &  -471.758 & * & 4.258 & 1 & 0.900 & 4 & 3.693 & 2 &       &   &       &   &      &   &      &  \\
\hline
 84 &  34 &  -727.341 & * &       &   &       &   &       &   &       &   &       &   & 1.33 & 9 &      &  \\
 86 &  36 &  -749.235 & 2 &       &   &       &   & 4.184 & 1 &       &   &       &   & 1.33 & 3 &      &  \\
 88 &  38 &  -768.467 & 1 & 4.994 & 1 & 0.923 & 1 & 4.220 & 1 &       &   &       &   & 1.30 & 2 &      &  \\
 90 &  40 &  -783.893 & 1 & 5.040 & 1 & 0.957 & 1 & 4.269 & 1 &       &   &       &   &      &   &      &  \\
 92 &  42 &  -796.508 & 1 & 5.104 & 1 & 0.950 & 1 & 4.315 & 1 &       &   &       &   & 1.40 & 2 &      &  \\
 94 &  44 &  -806.849 & 2 &       &   &       &   &       &   &       &   &       &   & 1.33 & 2 &      &  \\
 96 &  46 &  -815.034 & 2 &       &   &       &   &       &   &       &   &       &   &      &   &      &  \\
 98 &  48 &  -821.064 & 2 &       &   &       &   &       &   &       &   &       &   &      &   &      &  \\
\hline
134 &  52 & -1123.270 & 1 &       &   &       &   &       &   &       &   &       &   & 0.81 & 4 &      &  \\
136 &  54 & -1141.880 & 1 &       &   &       &   & 4.791 & 1 &       &   &       &   & 0.98 & 2 &      &  \\
138 &  56 & -1158.300 & 1 & 5.868 & 2 & 0.900 & 2 & 4.834 & 1 &       &   &       &   & 1.12 & 2 &      &  \\
140 &  58 & -1172.700 & 1 &       &   &       &   & 4.877 & 1 &       &   &       &   & 1.21 & 2 &      &  \\
142 &  60 & -1185.150 & 2 & 5.876 & 3 & 0.989 & 3 & 4.915 & 1 &       &   &       &   & 1.23 & 2 &      &  \\
144 &  62 & -1195.740 & 2 &       &   &       &   & 4.960 & 1 &       &   &       &   & 1.25 & 2 &      &  \\
146 &  64 & -1204.440 & 2 &       &   &       &   & 4.984 & 1 &       &   &       &   & 1.42 & 2 &      &  \\
148 &  66 & -1210.750 & 2 &       &   &       &   & 5.046 & 2 &       &   &       &   & 1.49 & 2 &      &  \\
150 &  68 & -1215.330 & 2 &       &   &       &   & 5.076 & 2 &       &   &       &   &      &   &      &  \\
152 &  70 & -1218.390 & 2 &       &   &       &   &       &   &       &   &       &   &      &   &      &  \\
\hline
206 &  80 & -1621.060 & 1 &       &   &       &   & 5.485 & 1 &       &   &       &   &      &   &      &  \\
210 &  84 & -1645.230 & 1 &       &   &       &   & 5.534 & 1 &       &   &       &   & 0.81 & 3 &      &  \\
212 &  86 & -1652.510 & 1 &       &   &       &   & 5.555 & 1 &       &   &       &   & 0.88 & 1 &      &  \\
214 &  88 & -1658.330 & 1 &       &   &       &   & 5.571 & 1 &       &   &       &   & 0.96 & 1 &      &  \\
216 &  90 & -1662.700 & 1 &       &   &       &   &       &   &       &   &       &   &      &   &      &  \\
218 &  92 & -1665.650 & 1 &       &   &       &   &       &   &       &   &       &   &      &   &      &  \\
\hline
\end{tabular}
\caption{\label{tab:fitdata2}
Experimental data for the fits, part II: along isotonic chains.
Doubly magic nuclei which would fit both sequences are not
repeated here.
For further explanations see table \ref{tab:fitdata1}
}
\end{table*}

\begin{table*}
\begin{tabular}{lrrrrrclrc}
\hline
 force & \multicolumn{1}{c}{$t_0$}
       & \multicolumn{1}{c}{$t_1$}
       & \multicolumn{1}{c}{$t_2$}
       & \multicolumn{1}{c}{$t_3$}
       & \multicolumn{1}{c}{$t_4$}
       &\hspace*{1em}
       & \multicolumn{1}{c}{$V_\mathrm{Pair,p}$}
       & \multicolumn{1}{c}{$V_\mathrm{Pair,n}$}
       & \multicolumn{1}{c}{$\rho_\mathrm{pair}$}
       \\   
       & \multicolumn{1}{c}{$x_0$}
       & \multicolumn{1}{c}{$x_1$}
       & \multicolumn{1}{c}{$x_2$}
       & \multicolumn{1}{c}{$x_3$}
       & \multicolumn{1}{c}{$b'_4$}
       && \multicolumn{1}{c}{$\alpha$}
       &
       & \multicolumn{1}{c}{$\eta_\mathrm{tls}$}
     \\
\hline
 SV-min & -2112.248&   295.781&   142.268& 13988.567&   111.291&&    601.160&   567.191&  0.21159\\
        &  0.243886& -1.434926& -2.625899&  0.258070&   45.93615&&  0.255368&&   0\\
\hline
\hline
 SV-bas & -1879.639&   313.749&   112.677& 12527.376&   124.634&&    674.618&   606.902&  0.20113\\
        &  0.258546& -0.381689& -2.823638&  0.123229&   34.11167&&     0.30& &  0\\
\hline
\hline
SV-K241 & -1745.184&   310.497&     5.705& 11975.552&   123.469&&    675.378&   610.425&  0.20108\\
        &  0.291787& -0.106990&-31.904438&  0.157335&   34.61687&&     0.34& &  0\\
\hline
SV-K226 & -2055.773&   317.043&   247.652& 13344.392&   126.375&&    619.478&   561.830&  0.21416\\
        &  0.217498& -0.717223& -1.975946&  0.081932&   33.12488&&     0.26& &  0\\
\hline
SV-K218 & -2295.822&   320.278&   330.798& 14557.194&   130.033&& 567.192&   517.838&  0.23131\\ 
        &  0.191803& -0.925134& -1.800814&  0.068066&   29.57382&&     0.22& &  0\\
\hline
\hline
SV-mas10 & -1813.907&   270.452&    57.215& 12965.454&   127.649&&    665.102&   588.101&  0.19774\\
         &  0.232898&  0.127181& -4.795203& -0.062346&   27.13922&&     0.33& &  0\\
\hline
SV-mas08 & -1982.650&   368.228&   274.611& 12141.094&   119.794&&    678.404&   621.138&  0.20781\\
         &  0.285653& -1.037404& -1.780825&  0.339364&   44.33569&&     0.26& &  0\\
\hline
SV-mas07 & -2203.658&   438.349&   566.845& 12222.736&   113.522&&    752.655&   688.246&  0.20122\\
         &  0.354369& -1.782940& -1.436932&  0.632534&   57.76849&&     0.20& &  0\\
\hline
\hline
SV-sym34 & -1887.367&   323.804&   351.782& 12597.277&   139.310&&    673.808&   603.478&  0.20008\\
         & -0.230110& -0.959586& -1.775475& -0.721854&   20.28327&&     0.30& &  0\\
\hline
SV-sym32 & -1883.278&   319.184&   197.329& 12559.469&   132.745&&    676.730&   607.540&  0.19995\\
         &  0.007688& -0.594307& -2.169215& -0.309537&   26.67638&&     0.30& &  0\\
\hline
SV-sym28 & -1877.431&   307.255&   140.868& 12511.940&   115.556&&    590.035&   539.892&  0.22395\\
         &  0.517821& -0.431291& -2.474137&  0.568794&   42.14865&&     0.30& &  0\\
\hline
\hline
SV-kap06 & -1880.594&   314.373&   194.940& 12537.049&   138.722&&    655.117&   575.327&  0.20717\\
         &  0.183805&  0.082542& -2.161972& -0.146674&   20.34984&&     0.30& &  0\\
\hline
SV-kap02 & -1878.883&   313.245&    44.042& 12519.929&   110.511&&    664.612&   616.996&  0.20136\\
         &  0.331398& -0.879672& -5.267351&  0.390236&   48.25944&&     0.30& &  0\\
\hline
 SV-kap00&-1877.891&   312.600&     7.104& 12509.993&    95.297& &   683.599&   648.780 & 0.19611\\
         & 0.393391& -1.454820&-26.089659&  0.640506&   63.35492& &  0.30 &&  0\\
\hline
\hline
SV-tls & -1879.892&   317.952&    30.265& 12531.858&   184.991& &   645.630&   577.394&  0.20723\\
       &  0.246413& -0.197627& -7.212765&  0.103793&    0.00010&&     0.30& &  1\\
\hline
\end{tabular}
\caption{\label{tab:forces}
The parameters of the energy-density functional (\ref{eq:enfun}) 
for the various SHF parameterizations used in  this paper. 
All parameterizations use a proton mass of 
$\hbar^2/{2m_p} = 20.749821$ and a neutron mass of
$\hbar^2/{2m_n} = 20.721260$.
For an explanation of the 
labels see table \ref{tab:nucmat}.
}
\end{table*}


\bibliographystyle{apsrev}
\bibliography{biblio,reviews,books,add}

\end{document}